\documentclass[structabstract]{aa}
\usepackage{graphicx}
\usepackage{longtable}
\usepackage[authoryear]{natbib}
\usepackage{txfonts}
\def\etal{\hbox{et al.}~}

\def\kms{\,km\,s$^{-1}$}

\def\hal{H$\alpha$}
\def\hbeta{H$\beta$}

\def\he{He\,{\sc i} 5876 \AA}
\def\hei{He\,{\sc i}}

\def\msun{{M$_{\odot}$}}
\def\rsun{{R$_{\odot}$}}

\def\mstar{{M$_{\star}$}}
\def\rstar{{R$_{\star}$}}

\begin{document}
   \title{Accretion dynamics in the classical T Tauri star V2129 Oph
          \thanks{This work uses observations made with the HARPS 
          instrument at the 3.6-m ESO telescope (La Silla, Chile) in the 
          framework of the LP182.D-0356.}}

   \author{S.H.P. Alencar\inst{1},
           J. Bouvier\inst{2},
           F.M. Walter\inst{3},
           C. Dougados\inst{2},
           J.-F. Donati\inst{4},
           R. Kurosawa\inst{5},
           M. Romanova\inst{5},
           X. Bonfils\inst{2},
           G.H.R.A. Lima\inst{1,2},
           S. Massaro\inst{6},
           M. Ibrahimov\inst{7},
          \and
           E. Poretti\inst{8}
          }

   \offprints{silvia@fisica.ufmg.br}

   \institute{Departamento de F\'{\i}sica -- ICEx -- UFMG, Av. Ant\^onio Carlos, 6627, 30270-901,
              Belo Horizonte, MG, Brazil
	 \and UJF-Grenoble 1 / CNRS-INSU, Institut de Plan\'etologie et d'Astrophysique de Grenoble (IPAG) UMR 5274, Grenoble, F-38041, France
         \and Department of Physics and Astronomy, Stony Brook University, Stony Brook, NY, 11794-3800, USA
         \and IRAP-UMR 5277, CNRS and Univ. de Toulouse, 14, Av. E. Belin, F-31400 Toulouse, France
	 \and Department of Astronomy, Cornell University, Space Sciences Building, Ithaca, NY 14853-6801, USA
         \and INAF - Osservatorio Astronomico di Palermo, Piazza del Parlamento 1, 90134 Palermo, Italy
         \and Ulugh Beg Astronomical Institute of the Uzbek Academy of Sciences, Astronomicheskaya 33, 700052 Tashkent, Uzbekistan
         \and INAF - Osservatorio Astronomico di Brera, via E. Bianchi 46, 23807 Merate (LC), Italy}

   \date{Received ; accepted }


\abstract
{Classical T Tauri stars are variable objects on several timescales,
but just a few of them have been studied in detail, with different observational
techniques and over many rotational cycles to enable the analysis of the 
stellar and circumstellar variations on rotational timescales.
}
{We test the dynamical predictions of the magnetospheric accretion model
with synoptic data of the classical T Tauri star V2129 Oph obtained over several
rotational cycles.
}
{We analyze high resolution observations obtained with the HARPS,
ESPaDOnS, and SMARTS spectrographs and simultaneous photometric
measurements, clearly sampling four rotational cycles, and fit them 
with cold/hot spot models and radiative transfer models
of emission lines.
}
{The photometric variability and the radial velocity variations in the photospheric lines 
can be explained by the rotational modulation due to cold spots, while the radial velocity variations of the 
He\,{\sc i} (5876 \AA) line and the veiling variability are due to hot spot rotational modulation.
The hot and cold spots are located at high latitudes and about the same phase, but
the hot spot is expected to sit at the chromospheric level, while the cold spot 
is at the photospheric level.
The mass accretion rate of the system is stable overall around $(1.5 \pm 0.6) \times 10^{-9}$
\msun yr$^{-1}$, but can increase by three times this value in a rotational cycle, during an accretion
burst. The \hal \ and \hbeta \ emission-line profiles vary substantially and are well-reproduced by radiative
transfer models calculated from the funnel flow structure of three-dimensional magnetohydrodynamics simulations, using the 
dipole+octupole magnetic-field configuration previously proposed for the system. 
Our diskwind models do not provide a significant contribution to the emission or absorption \hal \ line 
profile of V2129 Oph.
}
{The global scenario proposed by magnetospheric accretion for classical T Tauri stars
is able to reproduce the spectroscopic and photometric variability observed in V2129 Oph.
} 

   \keywords{Stars: pre-main sequence --
                Techniques: photometry, spectroscopy --
                Accretion, accretion disks
               }

\titlerunning{Accretion dynamics in the classical T Tauri star V2129 Oph}
\authorrunning{Alencar \etal }
\maketitle

\section{Introduction}\label{introduction}
\hspace{1.5em}
Classical T Tauri stars (CTTSs) are young, magnetically active low-mass ($M < 2$ \msun) 
stars that display signs of accretion from a circumstellar disk. They show
an emission excess with respect to the photosphere at wavelengths from X-rays to the radio
and are both spectroscopically and photometrically variable \citep{bou07}. The CTTSs have
strong permitted emission lines that vary on short timescales (days),
a redshifted absorption component associated with the accreting material,
and a blueshifted absorption component due to winds. They also show forbidden 
emission lines, which originate from a wind/jet that is powered by the accretion process. Magnetospheric
accretion models have been successfull in explaining most of the observed characteristics of CTTSs
\citep{shu94,har94,muz01,kur06,lim10} and are currently the most accepted ways to describe
these systems. In the past decade, magneto-hydrodynamic (MHD) simulations have 
followed the rotational evolution of CTTSs over as many as hundreds of rotational periods, presenting
a dynamical view of the star-disk interaction \citep{goo99,rom02,lon08,zan09} that can
be tested with synoptic observations.

Accretion has a long-lasting role in early stellar evolution, by providing
mass and helping to regulate the angular momentum transfer from the star to the
disk. The accretion process also affects the disk evolution, since the
high energy radiation produced in UV and X-rays influences the circumstellar disk 
lifetime \citep{gor09,owe10}. Furthermore, it has been suggested that accretion may affect
the early evolution of the star itself, by modifying the stellar radius at young ages \citep{bar09}. 
Therefore, it is important to understand the effects of the accretion process, 
and CTTSs are excellent laboratories for such a study.

V2129 Oph is a K5 CTTS \citep[$T_{\rm eff} \approx 4500$ K,][]{don07}
that has a maximum visual brightness of $m_V=11.2$ \citep{gra08}.
It is a young system (2-3 Myr) located in the $\rho$ Oph star-forming region
at $120 \pm 5$ pc \citep{loi08} and seen at 
moderate inclination with respect to our line of sight ($i\sim 60$\degr).
The measured values of rotational periods vary in the literature from
6.35 to 6.6 days \citep{gra08}, which is indicative of differential
rotation.
V2129 Oph has been the focus of a large observing campaign, including
simultaneous or quasi-simultaneous optical and near-IR photometry, high-resolution
spectroscopy and spectropolarimetry, and Chandra X-ray observations.
The spectropolarimetric data that we use in the present paper, which were obtained with ESPaDOnS, 
were previoulsy analyzed by \citet{don11}. These authors obtained the magnetic field structure
at the surface of the star for two observing epochs (in 2005 and 2009),
and found it is composed of a dipole and an octupole tilted by
about 20\degr\ with respect to the rotation axis. The magnetic field components
had intensities that varied between the two epochs, which implies that 
the magnetic field has a non-fossil origin. 
\citet{rom11} computed a numerical three-dimensional magnetohydrodynamics model of the 2005 
magnetic field configuration of V2129 Oph and simulated accretion onto the star.
They showed that the disk is truncated by the dipole component and that accretion
proceeds towards the star in two main accretion streams. Close to the stellar surface,
the flow is redirected by the octupolar component.
\citet{arg11} analyzed Chandra X-ray observations obtained in 27-29 June 2009,
simultaneously with the HARPS data analyzed in the present paper. They showed
that the soft X-ray emission observed in V2129 Oph, corresponding to dense cool plasma
of a few MK, comes from material heated in the accretion shock. The variability of
this soft X-ray emission in the observations is attributed to changes in the viewing 
angle of the accretion shock, as the system rotates. 

We present here an analysis of HARPS, ESPaDOnS, and SMARTS echelle spectra, obtained 
over a period of 73 days, together with simultaneous photometric measurements. 
Most of our spectra were taken from June 10 to July 13 2009,
providing a coverage of almost four rotational cycles. Although the ESPaDOnS data
were previously analyzed to determine the magnetic field configuration 
of the system by \citet{don11}, we use them here to complete our dataset in 
the time domain and to analyze the structure of the magnetospheric accretion flow. 
We describe the accretion diagnostics and
propose a viable scenario for the magnetospheric configuration and the 
star-disk interaction. We compare the observed Balmer emission lines 
with radiative transfer models calculated for the funnel flow structure 
obtained from 3D MHD simulations, using the 2009 magnetic field configuration.
We also calculate \hal \ line profiles with a hybrid magnetospheric and diskwind
model to investigate the importance of the wind to the observed line profiles.

\section{Observations}
\hspace{1.5em}
\subsection{Spectroscopy}
We present in Table \ref{tab_spectroscopy} a journal of spectroscopic observations.
Some of the spectroscopic observations of V2129 Oph were carried out from
June 10 to June 30, 2009, at ESO, La Silla with the HARPS dual fiber
echelle spectrograph \citep{may03}, covering the 3800 \AA\ to 6900 \AA\
spectral domain at a spectral resolution of $\lambda/\Delta\lambda 
\sim 80\ 000$. We acquired 28 high-resolution
spectra, over 20 non-consecutive nights at the 3.6 m telescope.
The data was automatically reduced by the HARPS Data Reduction Software.
The reduction procedure includes optimal extraction of the orders
and flat-fielding, wavelength calibration, and the removal of cosmic rays.
Spectropolarimetric observations were collected
from July 01 to July 14, 2009, with ESPaDOnS on CFHT.
The reduction of these ESPaDOnS data is described in detail in \citet{don11}.
Spectra were also obtained with the SMARTS 1.5m telescope at CTIO (Chile).
Two spectrographs were used. The low-to-moderate long-slit RC spectrograph was used
with the 300\arcsec\ long-slit, oriented east-west.
The detector was a Loral 1K CCD, with a 260 x 1199 readout region.
A total of 13 spectra were obtained from May 12 (JD 2454964.81) to July 7,
2009 (JD 2455020.73) with different settings, covering either the \hbeta \ or
\hal \ spectral regions.
We obtained three images at each epoch in order to filter cosmic rays.
Each set of images was accompanied by a wavelength calibration exposure of a
Ne-Ar or Th-Ar arc lamp. The image were bias-subtracted, trimmed, and flattened using
dome flats obtained each night. We co-added the three images using a median filter.
We also obtained echelle spectra with the 4.0 m Cassegrain echelle, now
bench-mounted and fiber-fed from the SMARTS 1.5 m telescope. 
A total of 19 spectra were obtained from May 1 (JD 2454953.93) to July 10,
2009 (JD 2455023.71).
A 100 micron slit was used, resulting in a resolution of about
30,000. The 31.6 l/mm echelle was used with a 226 l/mm cross-disperser and
the detector was a 2k SITe CCD with an ARCON controller.
Full wavelength coverage was achieved from 4020 \AA\ to 7300 \AA.
We obtained three 20 minute exposures at most epochs, and median-filtered the
data to reject cosmic rays. 
We located the orders and extracted the spectra by fitting a Gaussian
perpendicular to the dispersion direction at each pixel. 
We determined the precise wavelength solution by cross-correlating the Th-Ar
images obtained along with the spectra, with the template solution.
We corrected the velocities to the barycenter of the solar system.

\begin{table*}
\caption{Spectrograph, orbital phase, observing date, veiling, photospheric, and \hei \ radial velocities, \hal, \hbeta, and \hei \ equivalent widths and mass 
accretion rate}\label{tab_spectroscopy}
\begin{tabular}{llllllllll}
\hline \hline
Spectrograph & $JD-2\ 450\ 000$  & Phase & Veiling & $v_{\rm lsr}$ phot & $v_{\rm lsr}$ \hei & \hal \ eqw & \hbeta \ eqw & \hei \ eqw & $\dot{M}_{\rm acc}$ \\ 
             &                   &       &         & (\kms)             & (\kms)              & (\AA)     & (\AA)        & (\AA)      & ($10^{-9} \times$ \msun yr$^{-1}$)\\ 
\hline
HARPS        & 4993.52           & 0.59  & 0.23    & $-10.43$           & 2.49      & 16.88   & 4.19      & 0.49            & 4.36\\
HARPS        & 4993.61           & 0.61  & 0.19    & $-9.97$            & 2.69      & 15.03   & 3.86      & 0.41            & 3.45\\
HARPS        & 4993.71           & 0.62  & 0.19    & $-9.40$            & 3.56      & 13.80   & 3.62      & 0.42            & 3.64\\
SMARTS       & 4993.75           & 0.63  & --      & --                 & --        & 14.50   & --        &                 &     \\
HARPS        & 4994.57           & 0.75  & 0.07    & $-7.15$            & 3.21      & 8.87    & 1.87      & 0.18            & 1.08\\
HARPS        & 4994.65           & 0.76  & 0.06    & $-7.06$            & 2.39      & 8.87    & 1.63      & 0.19            & 1.17\\
HARPS        & 4994.76           & 0.78  & 0.07    & $-6.91$            & 2.09      & 8.95    & 1.25      & 0.18            & 1.10\\
HARPS        & 4995.56           & 0.90  & 0.04    & $-7.22$            & 0.30      & 12.12   & 1.60      & 0.12            & 0.680\\
HARPS        & 4995.65           & 0.92  & 0.06    & $-7.18$            & 1.89      & 14.25   & 1.83      & 0.14            & 0.865\\
HARPS        & 4995.74           & 0.93  & 0.10    & $-7.33$            & 2.54      & 15.13   & 2.49      & 0.16            & 1.07\\
HARPS        & 4996.56           & 1.06  & 0.07    & $-6.83$            & 1.62      & 13.85   & 2.41      & 0.15            & 1.13\\
HARPS        & 4996.59           & 1.06  & --      & --                 & --        & 13.52   & 2.38      & 0.16            & 1.24\\
HARPS        & 4998.54           & 1.36  & 0.10    & $-4.72$            & 0.87      & 31.31   & 5.06      & 0.29            & 2.28\\
HARPS        & 4998.58           & 1.37  & 0.10    & $-4.76$            & $-4.70$   & 30.48   & 5.53      & 0.27            & 2.05\\
HARPS        & 4998.76           & 1.39  & 0.16    & $-5.34$            & $-4.91$   & 27.34   & 4.95      & 0.28            & 2.08\\
HARPS        & 4999.54           & 1.51  & 0.28    & $-7.50$            & $-3.36$   & 16.72   & 3.88      & 0.48            & 3.88\\
HARPS        & 4999.67           & 1.53  & 0.30    & $-9.14$            & 2.39      & 17.52   & 4.27      & 0.51            & 4.44\\
HARPS        & 5001.61           & 1.83  & 0.14    & $-7.20$            & 3.16      & 12.83   & 1.83      & 0.25            & 1.81\\
HARPS        & 5001.69           & 1.84  & 0.14    & $-7.21$            & 3.01      & 12.58   & 1.65      & 0.24            & 1.71\\
HARPS        & 5002.57           & 1.98  & 0.13    & $-7.26$            & 3.26      & 15.59   & 3.43      & 0.24            & 2.03\\
HARPS        & 5002.60           & 1.98  & --      & --                 & --        & 14.98   & 2.86      & 0.19            & 1.47\\
HARPS        & 5002.68           & 1.99  & 0.11    & $-6.99$            & 2.34      & 15.05   & 3.36      & 0.25            & 2.22\\
HARPS        & 5003.50           & 2.12  & 0.07    & $-6.56$            & 2.59      & 18.30   & 3.66       & 0.20            & 1.61\\
HARPS        & 5004.52           & 2.28  & 0.08    & $-5.43$            & 0.00      & 15.94   & 3.33      & 0.21            & 1.55\\
SMARTS       & 5004.77           & 2.31  & --      & --                 & --        & 15.13   & --        &                 &     \\
SMARTS       & 5005.68           & 2.45  & --      & --                 & --        & 15.61   & --        &                 &     \\
HARPS        & 5006.50           & 2.58  & --      & --                 & --        & 10.70   & 2.01      & 0.27            & 1.87\\
SMARTS       & 5006.67           & 2.60  & --      & --                 & --        & 14.16   & --        &                 &     \\
HARPS        & 5007.51           & 2.73  & 0.11    & $-7.34$            & 0.30      & 11.32   & 2.41      & 0.28            & 2.08\\
HARPS        & 5008.52           & 2.89  & 0.10    & $-8.16$            & $-2.82$   & 19.26   & 2.88      & 0.19            & 1.27\\
SMARTS       & 5009.73           & 3.07  & --      & --                 & --        & 18.40   & --        &                 &     \\
HARPS        & 5010.50           & 3.19  & 0.11    & $-6.14$            & 1.02      & 15.74   & 3.34      & 0.18            & 1.31\\
SMARTS       & 5010.74           & 3.23  & --      & --                 & --        & 16.38   & --        &                 &     \\
HARPS        & 5012.49           & 3.50  & 0.19    & $-7.77$            & 2.09      & 12.91   & 2.66      & 0.31            & 2.00\\
SMARTS       & 5012.74           & 3.53  & --      & --                 & --        & 13.89   & --        &                 &     \\
SMARTS       & 5013.73           & 3.69  & --      & --                 & --        & 14.49   & --        &                 &     \\
EsPADoNS     & 5013.77           & 3.69  & 0.08    & $-7.85$            & 2.89      & 14.30   & 2.49      & 0.27            & 2.00\\
EsPADoNS     & 5013.90           & 3.71  & 0.08    & $-8.56$            & 3.81      & 13.27   & 2.66      & 0.27            & 2.07\\
SMARTS       & 5014.56           & 3.81  & --      & --                 & --        & 18.49   & --        &                 &     \\
EsPADoNS     & 5014.77           & 3.85  & 0.07    & $-7.92$            & 1.22      & 18.07   & 3.13      & 0.16            & 0.960\\
EsPADoNS     & 5014.90           & 3.87  & 0.06    & $-8.72$            & 1.07      & 18.16   & 3.21      & 0.17            & 1.06\\
SMARTS       & 5015.75           & 3.99  & --      & --                 & --        & 17.09   & --        &                 &     \\
EsPADoNS     & 5015.76           & 4.00  & 0.05    & $-8.28$            & 1.42      & 19.52   & 4.25      & 0.18            & 1.37\\
EsPADoNS     & 5015.89           & 4.02  & 0.05    & $-7.32$            & $-0.82$   & 19.12   & 3.60      & 0.13            & 0.966\\
EsPADoNS     & 5017.76           & 4.30  & 0.11    & $-6.20$            & $-5.01$   & 20.88   & 2.92      & 0.22            & 1.64\\
EsPADoNS     & 5017.88           & 4.32  & 0.11    & $-5.46$            & $-3.93$   & 19.73   & 3.47      & 0.28            & 1.62\\
EsPADoNS     & 5018.76           & 4.46  & 0.06    & $-7.46$            & $-2.77$   & 14.80   & 3.89      & 0.28            & 1.95\\
EsPADoNS     & 5018.88           & 4.48  & 0.04    & $-8.07$            & $-2.24$   & 14.46   & 3.39      & 0.23            & 1.37\\
SMARTS       & 5019.60           & 4.59  & --      & --                 & --        & 14.94   & --        &                 &     \\
EsPADoNS     & 5019.76           & 4.61  & 0.05    & $-9.54$            & $-0.42$   & 14.41   & 2.38      & 0.19            & 1.15\\
EsPADoNS     & 5019.89           & 4.63  & 0.05    & $-8.89$            & 0.30      & 14.52   & 2.81      & 0.20            & 1.26\\
EsPADoNS     & 5020.76           & 4.76  & 0.07    & $-7.62$            & 2.39      & 14.21   & 2.04      & 0.26            & 1.82\\
EsPADoNS     & 5020.88           & 4.78  & 0.06    & $-7.03$            & 2.64      & 16.37   & 2.23      & 0.26            & 1.84\\
EsPADoNS     & 5021.76           & 4.92  & 0.04    & $-7.60$            & 0.97      & 15.30   & 2.50      & 0.14            & 0.847\\
EsPADoNS     & 5021.88           & 4.93  & 0.04    & $-8.45$            & 1.07      & 16.34   & 3.10      & 0.14            & 0.867\\
EsPADoNS     & 5022.75           & 5.07  & 0.06    & $-6.96$            & 1.99      & 16.73   & 2.75      & 0.12            & 0.822\\
EsPADoNS     & 5022.88           & 5.09  & 0.06    & $-8.00$            & 0.30      & 16.10   & 3.01      & 0.12            & 0.811\\
SMARTS       & 5023.71           & 5.21  & --      & --                 & --        & 31.40   & --        &                 &     \\
EsPADoNS     & 5024.78           & 5.38  & 0.07    & $-6.33$            & $-2.19$   & 20.31   & 3.06      & 0.29            & 2.27\\
EsPADoNS     & 5025.85           & 5.54  & 0.07    & $-9.88$            & 0.25      & 15.70   & 5.14      & 0.33            & 2.38\\
EsPADoNS     & 5026.85           & 5.69  & 0.06    & $-8.35$            & 2.49      & 11.05   & 4.12      & 0.27            & 2.01\\
\hline
\end{tabular}
\end{table*}

\subsection{Photometry}
The journal of photometric observations is presented in Table \ref{tab_photometry}.
Our $BVRIJHK$ photometric observations were carried out with 
the ANDICAM dual-channel imager at the SMARTS 1.3 m telescope at Cerro Tololo. 
The optical data is composed of 10 sets of observations on 10 nights, between
23 June (JD 2455007.57) to 8 July 2009 (JD 2455021.64). 
BKLT J162749-242540 (ROC 31) was used as a comparison star.
The near-IR data consists of 13 sets of observations on 13 nights, between 
21 June (JD 2455005.65) and 8 July 2009 (JD 2455021.65).
Integration times were 4 seconds (the minimum allowed). We obtained 3 images
through each filter at different dither positions that were shifted and co-added.
Measurements are differential with respect to the star
BKLT J162746-242323 (Elias 2-35; VSSG 13), which was a highly reddened background
K5III star viewed through the dark cloud. Published 2MASS magnitudes of the 
comparison star were used to convert the instrumental magnitudes to apparent magnitudes.

Seven sets of $V$ and $R$ observations on 7 nights were obtained at Palermo in 
the period between June 22 (JD 2455006.40) and June 29 (JD 2455013.40).
SR21 was used as a comparison star.
We also obtained 15 $UBVRI$ observations at Mount Maidanak in 7 nights
from 19 June (JD 2455002.29) to 30 June 2009 (JD 2455013.24). 

\begin{table}
\caption{Journal of photometric observations.}\label{tab_photometry}
\begin{tabular}{lllll}
\hline \hline
Observatory & $JD-2\ 450\ 000$  & Filters & Observer & $N_{obs}$ \\ 
\hline
CTIO        & 5007.57-5021.64 & $BVRI$  & F. Walter  & 10\\
CTIO        & 5005.65-5021.65 & $JHK$   & F. Walter  & 13\\
Palermo     & 5006.40-5013.40 & $VR$    & S. Massaro & 7\\
Maidanak    & 5002.29-5013.24 & $UBVRI$ & M. Ibrahimov & 15\\
\hline
\end{tabular}
\end{table}


\section{Results}
\hspace{1.5em}
In our analysis, we use the ephemeris published by \citet{don07} for
V2129 Oph in the phase calculations
\begin{equation}
{\rm HJD} = 245\,3540.0 + 6.53E,
\end{equation}
where 6.53 days is the adopted period of V2129 Oph.

\subsection{Photometric and spectral variations}\label{phot_spec_variations}
\begin{figure}[htb]
\centering
\includegraphics[width=9cm]{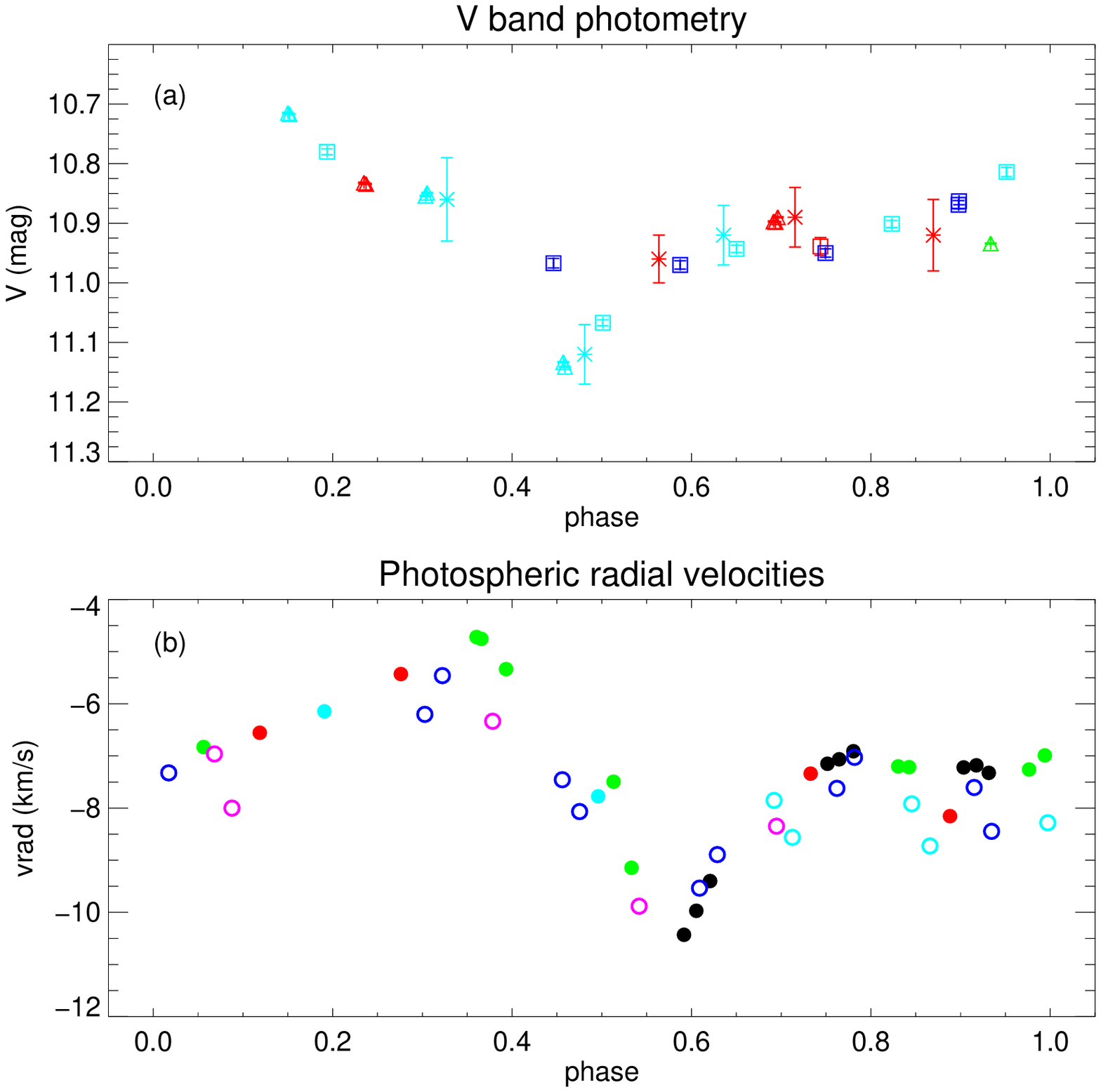}
\includegraphics[width=9cm]{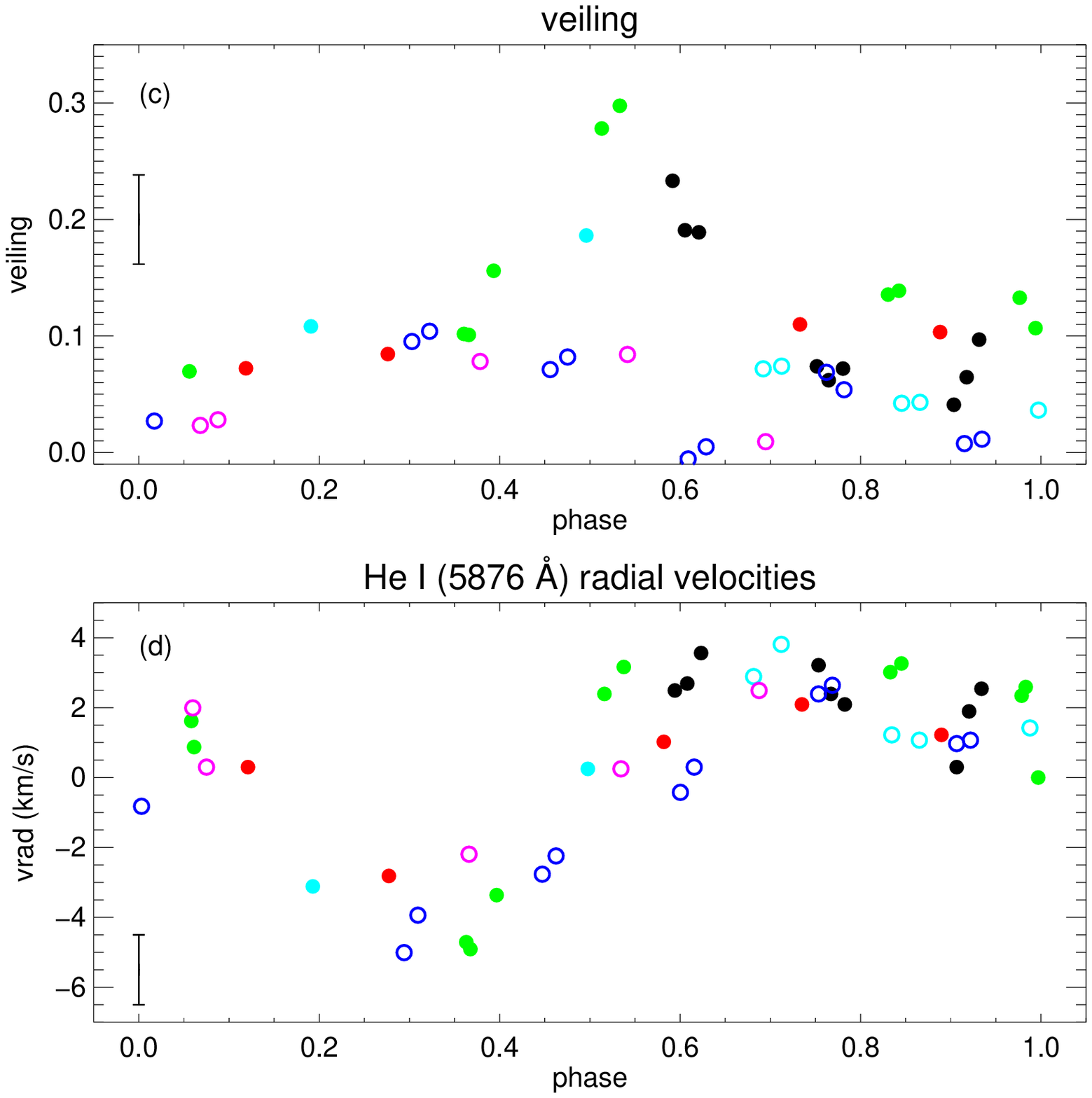}
\caption{V band photometry corrected from extinction (panel {\it a}).
SMARTS data are shown as squares,
Palermo data as stars, and Maidanak data as triangles. 
Photospheric radial velocities (panel {\it b}), 
veiling (panel {\it c}), and \hei \ (5876 \AA) radial velocity (panel {\it d})
phase variations.
In panels {\it b}, {\it c}, and {\it d}, filled symbols are from HARPS 
data and open ones from ESPaDOnS.
The error bars are shown for each observation in panel a, and are
smaller than the symbol sizes in panel b. We show in panels c and d
only the mean error bars to avoid crowding the figures. Different 
colors represent
different rotational cycles identified as the integer part
of the observational phase in Table \ref{tab_spectroscopy}:
black - cycle 0, green - cycle 1, red - cycle 2, aqua - cycle 3,
dark blue - cycle 4, pink - cycle 5.
}
\label{veiling_heIvrad}
\end{figure}

The photometric data show rotational modulation in all filters ($BVRI$), with 
an amplitude decreasing from the $B$ to the $I$ band. 
The $V$ data are presented in Fig. \ref{veiling_heIvrad} (panel a), after
correction for the interstellar extinction with $A_V=0.6$ \citep{don07}.

Photospheric radial velocities were measured on HARPS and ESPaDOnS 
spectra using the spectral region $5415$ \AA $< \lambda < 5465$ \AA, cross-correlating 
V2129 Oph spectra with spectra of the K7 weak T Tauri star (WTTS) V819 Tau 
(Fig. \ref{veiling_heIvrad}, panel b). 
The measured values from both spectrographs 
are in excellent agreement, and the values obtained for the HARPS spectra agree
with those directly provided by the HARPS pipeline by 
cross-correlating the object's spectrum with a K5 mask. 
The error in the HARPS and ESPaDOnS photospheric radial velocities is around 
0.2 \kms \ and 0.1 \kms, respectively, obtained by fitting the cross-correlation
and taking into account the signal-to-noise ratio (S/N) of the spectra. However, looking at Fig.
\ref{veiling_heIvrad} (panel b) we can see that the HARPS data (filled symbols)
are clearly less scattered than the ESPaDOnS ones (open symbols), which is probably due to 
the higher stability of the HARPS spectrograph.

\begin{figure}[htb]
\centering
\includegraphics[width=9cm]{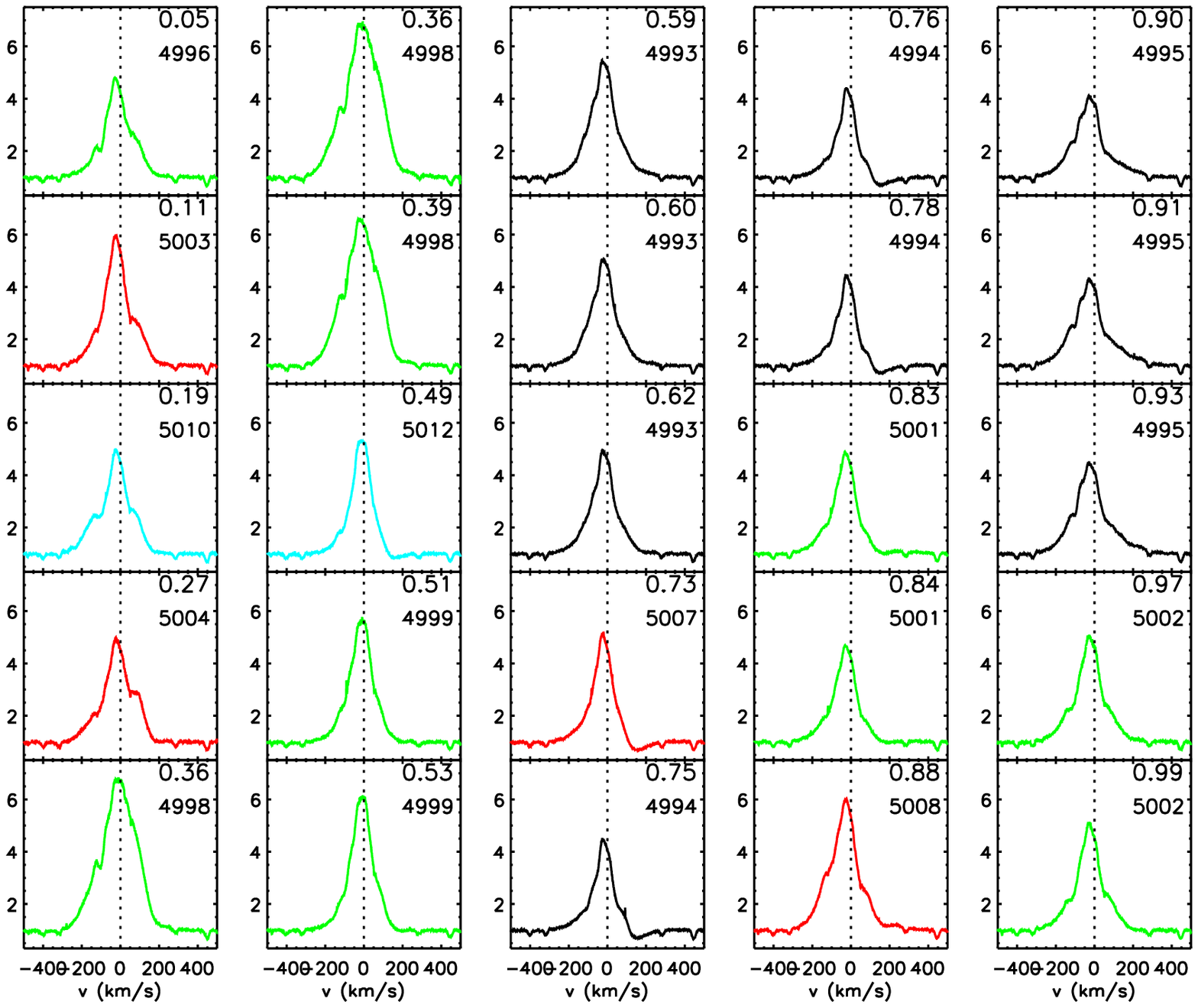}
\caption{\hal\ profiles obtained with the HARPS spectrograph. 
The profiles are normalized to the continuum level and the vertical dotted lines
indicate the stellar rest velocity. Phases and JDs are given for each 
spectrum in the panels. The color code
is the same as in Fig. \ref{veiling_heIvrad}.
}
\label{halpha_harps}
\end{figure}

\begin{figure}[htb]
\centering
\includegraphics[width=9cm]{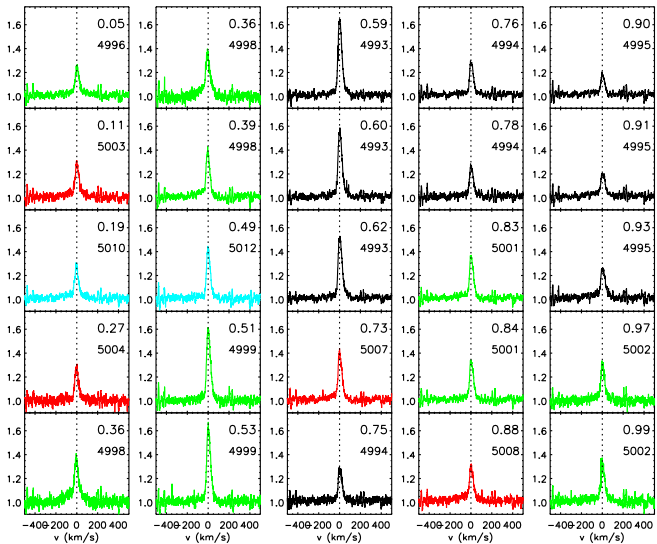}
\caption{\he\ profiles obtained with the HARPS spectrograph.
The profiles are normalized to the continuum level and the vertical dotted lines
indicate the stellar rest velocity. Phases and
JDs are given for each spectrum in the panels. The color code
is the same as in Fig. \ref{veiling_heIvrad}.
}
\label{heI_harps}
\end{figure}

We present in Figs. \ref{halpha_harps} and \ref{heI_harps} the \hal\ and
\hei \ line profiles observed with HARPS, and folded in phase.
The \hal\ line profile varies substantially in phase,
having the same characteristics in similar phases, such as
a red emission shoulder from phase 0.05 to 0.35 and a 
redshifted absorption component from phase 0.6 to 0.8.

The \hei \ (5876 \AA) line contains only a narrow component in the HARPS and ESPaDOnS spectra,
which we fitted with a single Gaussian to measure the emission line parameters.
The Gaussian central wavelength position was used to measure
the \hei \ radial velocity, which is clearly modulated over the rotational
phase (Fig. \ref{veiling_heIvrad}, panel d). The HARPS and ESPaDOnS measurements
are consistent and stable over a $\sim 33$-day timescale.
The error in the measured radial velocities is on the order of 0.5 to 2 \kms\ and
dominated by our fitting uncertainty in the emission line profile.

\subsection{Veiling}\label{veiling}
We measured veiling values in the spectral region $5415$ \AA $< \lambda < 5465$ \AA.
This region contains data with a good S/N and many photospheric lines that are not blended, and
is also devoid of emission lines.
We used the projected rotational velocity of V2129 Oph obtained by \citet{don07},
$v\sin{i}=14.5 \pm 0.3$ \kms, which agrees with our determination.
We initially used V819 Tau (K7) as a comparison star, but owing to its
spectral type mismatch with V2129 Oph, we checked our measured veiling values
with the WTTS V410 Tau (K5).

While V819 Tau rotates slowly \citep[$v \sin{i} < 15$ \kms,][]{her88}
and is an excellent standard star to use
in the determination of both $v\sin{i}$ and veiling, V410 Tau
is a rapid rotator \citep[$v\sin{i}=74\pm3$ \kms,][]{ske10}. We first
calculated the veiling of V410 Tau
by using V819 Tau as a standard, obtaining the value of $0.11 \pm 0.04$. This is not a true
veiling, since V410 Tau is not accreting, but it gives us an idea of the
veiling value created by the mismatch of spectral types between a K5 star
and a K7 standard. We then rotationally broadened the spectra of V2129 Oph
to the rotation velocity of V410 Tau and calculated the veiling of V2129 Oph
with V410 Tau as a standard star. The values obtained are consistent with
those determined using V819 Tau as a standard, but shifted to lower values
by $0.15 \pm 0.04$, which is consistent with the veiling induced by the
spectral type mismatch between V819 Tau and V2129 Oph.

The veiling error obtained
using V819 Tau as a standard is on the order of 0.01 and 0.02 for the HARPS and
ESPaDOnS data, respectively, while it reaches 0.04 when we use V410 Tau, 
owing to its high rotational velocity. We therefore decided to use the veiling
values obtained with V819 Tau that had been decreased by a factor of 0.15 to take into account the
spectral type difference between V819 Tau and V2129 Oph. Although the absolute
values still have an error of about 0.04 due to the scaling process, the relative
error between each measurement is on the order of 0.02.
The final result is presented in Fig. \ref{veiling_heIvrad} (panel c).
The veiling that we determined for the ESPaDOnS data are compatible, within the errors,
with the values calculated by \citet{don11}.

As shown in Sect. \ref{spots}, a large cold spot is actually located in the stellar photosphere,
which should affect the measured veiling values, as the system's continuum flux
is modulated by the spot. For $\Delta(V) \sim 0.21$ mag, the relative veiling
variation amounts to about 20\%, which is, however, much lower than the veiling variations 
we observe.

\subsection{Mass accretion rate}\label{mass accretion rate}
\begin{figure}[htb]
\centering
\includegraphics[width=9cm]{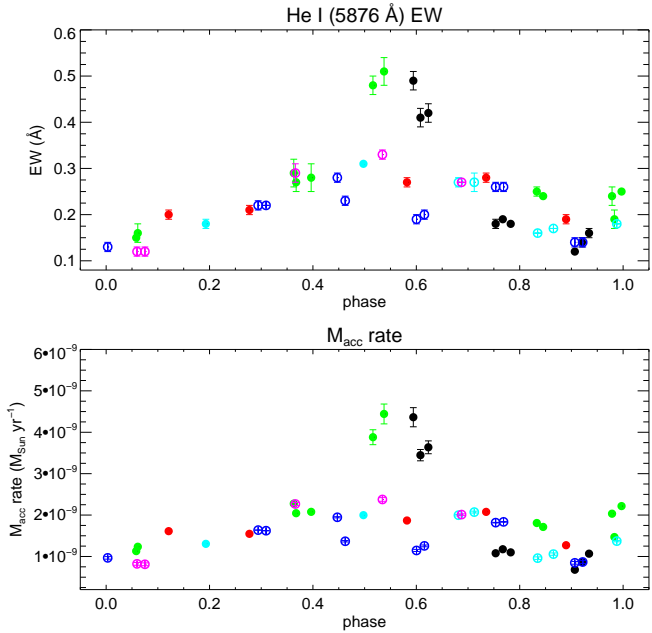}
\caption{Rotational variation in the \hei \ EW  and the mass accretion rate.
The mass accretion rate errors come from the relation between the \hei \ (5876\AA) luminosity
and the accretion luminosity by \citet{fan09}.
Filled symbols are from HARPS data and open ones from ESPaDOnS.
Different colors represent
different rotational cycles identified as the integer part
of the observational phase in Table \ref{tab_spectroscopy}:
black - cycle 0, green - cycle 1, red - cycle 2, aqua - cycle 3,
dark blue - cycle 4, pink - cycle 5.
}
\label{macc}
\end{figure}

We measured the \he\ line equivalent widths (EWs) (see Tab. \ref{tab_spectroscopy}
and Fig. \ref{macc}, top panel) 
and computed line fluxes using the $V$ band photometry. We do not have
simultaneous $V$ measurements for all the spectroscopic data, but the phase coverage
of the $V$ data is quite complete (Fig. \ref{veiling_heIvrad}, panel a). We therefore interpolated
the $V$ magnitude folded in phase at the rotational phase of the spectroscopic
observations. Using the empirical correlations of \citet{fan09}, we calculated the mass-accretion
rates values from the \hei\ line fluxes. The results are shown in Fig. 
\ref{macc}, where we see that the mass accretion rate remains fairly stable most of
the time at a value of $(1.5\pm 0.6)\times 10^{-9}$ \msun${\rm yr}^{-1}$, which agrees
with the mass accretion rate determined by \citet{don11}. 
However, we note that, around phase 0.5, the mass accretion rate varied 
substantially over a period of days during the HARPS observations. This phase corresponds 
to the main accretion spot facing the observer. 
There is further evidence from the \hal\ and \hbeta\ emission line profiles that the mass accretion
increased during the HARPS observations (see Sect \ref{emission}).

\section{Discussion}

\subsection{Hot and cold spots}\label{spots}
\begin{figure}[htb]
\centering
\includegraphics[width=9cm]{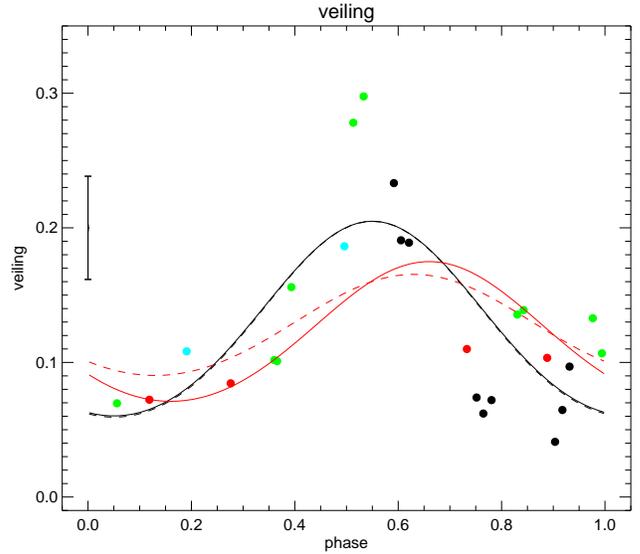}
\caption{Best-fit hot spot model for the HARPS veiling variations (black lines).
The red lines represent the veiling generated by the optimal hot spot
fits to the \hei \ radial velocity variations (Fig. \ref{heI_vrad_fit}).
Models with $i=60$\degr\ and $i=45$\degr\ are shown as solid and dashed lines, respectively .
The mean error bar in the observed veiling is shown on the left and the 
color code of the points is the same as in Fig. \ref{veiling_heIvrad}.
}
\label{veiling_fit}
\end{figure}

\begin{figure}[htb]
\centering
\includegraphics[width=9cm]{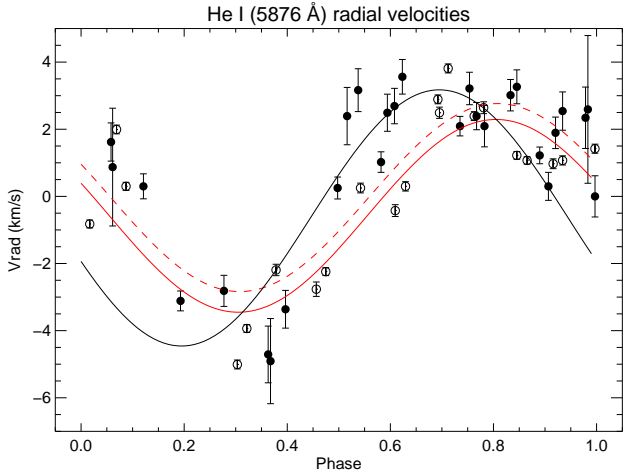}
\caption{HARPS (filled symbols) and ESPaDOnS (open symbols) 
\hei\ radial velocity measurements, with respective error bars.
The red curves represent the best-fit hot-spot model for the \hei\ radial 
velocity data obtained with $i=60$\degr \ (solid line) and $i=45$\degr \ 
(dashed line). The black solid line corresponds to the best fit
to the veiling variations with $i=60$\degr, 
presented in Fig. \ref{veiling_fit}.
}
\label{heI_vrad_fit}
\end{figure}

The veiling clearly varies in phase during the HARPS observations (Fig. \ref{veiling_heIvrad},
panel c), with a maximum
value of $\sim 0.30$ close to phase $\phi=0.5$ and a minimum value generally above zero, suggesting 
that the hot spot remains in view during the whole rotational cycle. This is confirmed
by the analysis of line profile variations (see Sect. \ref{models}). We developed a hot spot
model with a circular spot to reproduce the veiling modulation. 
We used $T_{\rm eff}=4500$ K, $T_{\rm spot}=8000$ K and varied the spot radius, 
latitude, and location with rotational phase. We run models with two different
values of the inclination of the system with respect to our line of sight, $i=45$\degr \
and $i=60$\degr, which correspond to values suggested in the literature for the system
\citep{don11}.
For a system inclination of $i=60$\degr, 
the best-fit model corresponds to a hot spot with a filling factor of 
1.2\%, located at phase 0.55 and high latitude (78\degr). If we use, instead, 
a system inclination of $i=45$\degr\, we obtain a hot spot with a filling factor of 0.3\%, 
located at phase 0.55 and latitude of 70.5\degr. The two solutions, 
which are indicated as black lines in Fig. \ref{veiling_fit}, are degenerate, having the 
same $\chi^2$ and, although they do not constrain 
the system inclination, they both converge to a major hot spot with very similar 
characteristics. The low observed veiling contrast during the rotational cycle can only be 
reproduced with a very high-latitude spot, i.e.,
close to the rotational axis. However, none of the models perfectly
reproduce the narrow veiling peak near phase 0.5, but instead predict a
much smoother modulation. We have overplotted in red in Fig. \ref{veiling_fit} the veiling
variations generated by the hot spot models that most closely reproduce the 
observed \he \ radial velocity curve (Fig. \ref{heI_vrad_fit}). The hot spot inferred from the 
\hei \ radial velocity variations is consistent overall with the observed
veiling changes.  We also note that the veiling 
variability was not seen in all the observed rotational cycles (Fig. \ref{veiling_heIvrad}), 
which suggests that a hot spot has evolved and dimmed on a timescale of about a week.
Very low veiling values are indeed derived from the ESPaDOnS spectra, which were acquired the week
after the HARPS measurements.

The \he\ line is thought to be produced in the post-shock region of the
accretion shock. The modulation of its radial velocity 
(Fig. \ref{veiling_heIvrad}, panel d) is then likely due to the hot 
spot rotation. The phase of the \hei\ radial velocity curve is indeed
exactly the opposite of that of
the photospheric radial velocity curve, the bluest velocity being reached at
$\phi\sim 0.35$ and the reddest at $\phi\sim 0.6$.
We assumed that all the \hei\ line is produced at the location of the hot spot
and fitted the measured radial velocities of \hei\ with a hot spot model.
We set the spot latitude and phase, the infall velocity, and the system 
inclination as free parameters. However, the system inclination was not well-constrained in
the fitting procedure, so we decided to compute models with inclination values
fixed at $i=45$\degr\ and $i=60$\degr, as we did in the veiling model calculations.
The best-fit results are shown in Table \ref{fit_vrad_he} and in
red in Fig. \ref{heI_vrad_fit}. The final values for each parameter obtained with $i=45$\degr \ and
$i=60$\degr \ agree within the errors and correspond to a hot spot located
at high latitude (80\degr) near phase 0.6 with an infall velocity of about 10 \kms.
This is consistent with the assumption that the hot spot
required to account for the veiling variations is also responsible for the
modulation of the \hei\ radial velocity. To illustrate this, we overplotted
in Fig. \ref{heI_vrad_fit}, in black, the \hei \ radial velocity model that corresponds
to the best fit of the veiling curve. That the \hei \ radial
velocity changes during the whole rotational cycle indicates that the hot spot 
is always in view, which agrees with the veiling analysis.
The small difference in spot parameters required to explain the veiling
variations and the \hei\ radial velocity curve are probably related to 
the basic assumption of circular spots at the stellar surface, an assumption 
that is probably wrong, but it would be beyond the scope of this paper to 
investigate more complex spot geometries.

\begin{table}
\caption{System inclination, hot spot latitude, phase, and infall velocity of
the best fits to the \hei\ radial velocity.} \label{fit_vrad_he}
\begin{tabular}{llll}
\hline \hline
i(\degr) & $lat$(\degr) & phase & $v_{inf}$(\kms) \\
\hline
45 & $80 \pm 6$ & $0.63 \pm 0.10$  & $10 \pm 2$\\
60 & $81 \pm 5$ & $0.66 \pm 0.10$  & $13 \pm 2$\\
\hline
\end{tabular}
\end{table}

\begin{figure}[htb]
\centering
\includegraphics[width=9cm]{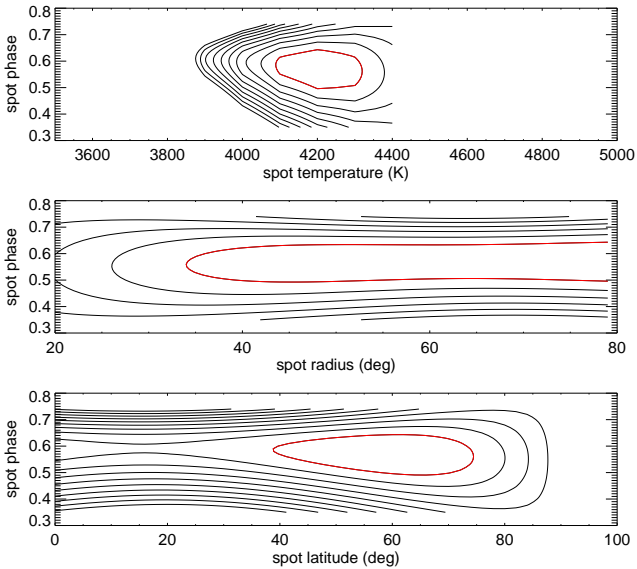}
\caption{$\chi^2$ contours of the cold spot models with $i=45$\degr.
Each contour level corresponds to an increase of $\chi^2$ minimum. 
The red lines correspond to a 2$\chi^2$ minimum
and represent the error in each parameter quoted in the text.
}
\label{chi2}
\end{figure}

\begin{figure}[htb]
\centering
\includegraphics[width=9cm]{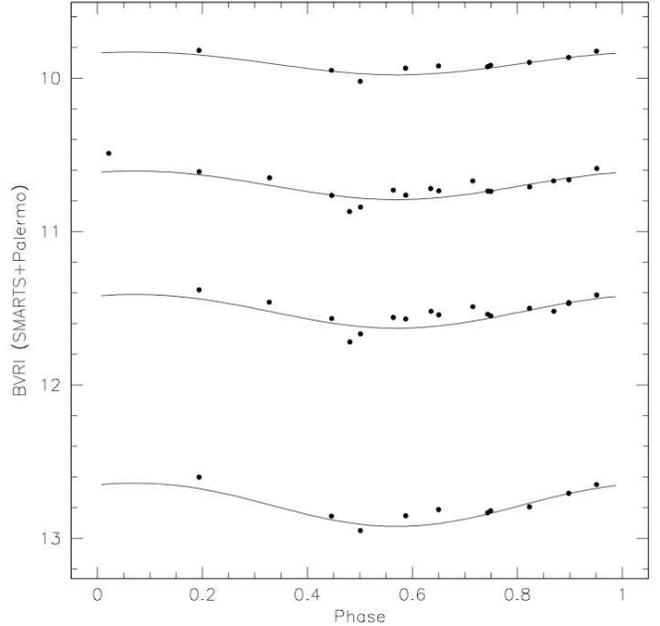}
\caption{Best-fit cold-spot model with $i=45$\degr\ to the photometric observations.
The parameters of the cold-spot model used to generate the above light curves 
correspond to a $\chi^2$ minimum ($T_{\rm spot}=4200$ K, spot latitude of 60\degr,
spot phase of 0.57, and spot filling factor of 40\%).
}
\label{spot_photometry}
\end{figure}

The photometric variations appear to be dominated by a low-level
modulation from a cool spot (Fig. \ref{veiling_heIvrad}, panel a) and 
the smooth shape of the light curve suggests that a
high latitude spot is always in view. A cold spot model was run with a fixed
inclination at either $i=45$\degr\ or $i=60$\degr\ to attempt
to reproduce the observed light curves, but the model solutions
are unable to efficently constrain all the free parameters, owing to the small number 
of data points available. The best-fit models with $i=45$\degr\ correspond to a cold
spot with temperature $T_{\rm spot}=(4200 \pm 100)$ K located at phase 
$0.57 \pm 0.07$ and latitude of 60\degr $\pm$ 20\degr. Unfortunately, the cold spot radius 
is almost unconstrained ($R_{\rm spot} > 35$\degr), which corresponds to a large incertitude
in the spot filling factor ($f > 9$\%).
We show in Fig. \ref{chi2} the $\chi^2$ contours of the
parameters adjusted in the cold spot models. In Fig. \ref{spot_photometry},
we present the fits to the photometric data of the best-fit cold spot model
with $i=45$\degr. 
The best-fit models with $i=60$\degr\ correspond to a cold
spot with temperature $T_{\rm spot}=(3500 \pm 500)$ K located at phase 
$0.57 \pm 0.07$ and latitude of 84\degr $\pm$ 5\degr. The cold spot radius 
is again unconstrained ($R_{\rm spot} > 50$\degr, $f > 18$\%), although
a large spot is clearly needed in both models.

\begin{figure}[htb]
\centering
\includegraphics[width=9cm]{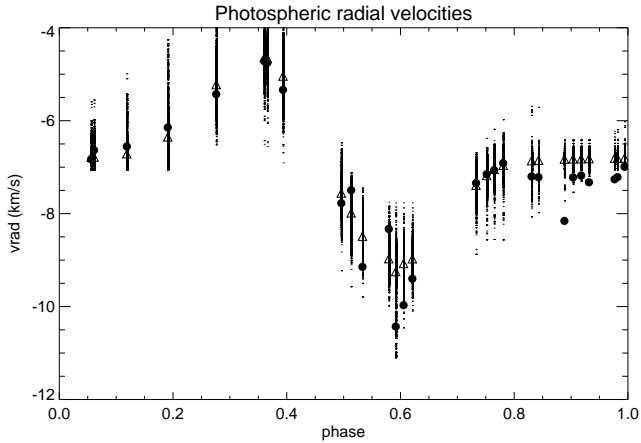}
\caption{Radial velocity variations induced by a cold spot model
obtained in 10000 Monte Carlo runs varying the parameter
space of possible solutions (small black dots). The triangles
represent the median values of the theoretical solutions
and the HARPS photospheric radial velocities are shown as
large filled circles.
}
\label{spot_spectro}
\end{figure}

As pointed out by \citet{pet11}, hot and cold spots
can distort in similar ways the photospheric line profiles, and photometry
taken simultaneously with the spectra is needed to distinguish between
the two effects. In the case of V2129 Oph, the photometric variations
clearly indicate that there is a large cold spot in the stellar photosphere
that would cause the photospheric radial velocity modulation 
seen in Fig. \ref{veiling_heIvrad} (panel b). 
Moreover, the photospheric radial velocity variations are very stable over 
the observed period and clearly present in all rotational cycles, including 
those with very low veiling values (Fig. \ref{veiling_heIvrad}, panel b). 
This also points to a cold spot as the most likely
cause of the photospheric radial velocity variations, instead of a hot spot.
The reddest velocity
is seen at $\phi\sim 0.35$, the bluest at $\phi\sim 0.6$,
and the mean velocity of $\sim -7$ \kms\ is crossed at $\phi\sim 0.5$,
indicating that the spot then faces the observer.
This is consistent with the assumption that
the cold spot required to account for the $BVRI$ light curves is also
responsible for the photospheric radial velocity modulation.
We then tried to reproduce the observed HARPS photospheric radial-velocity variations using cold
spot models with a system inclination of 60\degr. We varied the spot brightness, 
its latitude, the phase at which it occurs, and the systemic velocity of the star,
but we could not find a unique solution to the data.
We calculated 10000 solutions exploring the parameter space of possible
solutions and obtained probability density functions
for each parameter. Some parameters are well-constrained, such as
the phase where the spot faces
the observer ($\phi=0.475 \pm 0.025$), while others are clearly poorly constrained,
like the spot filling factor (9\% $< f <$ 50\%), its brightness ($0.0 <$ brightness $< 0.8$),
and the spot latitude ($0$\degr $<$ latitude $< 55$\degr). The solution is degenerated
for several combinations of the poorly constrained parameters. We present in
Fig. \ref{spot_spectro} the radial velocity variations due to a cold spot
calculated in each of the 10000 models run
at the phases of the HARPS observations and a system inclination of $i=60$\degr.

The hot and cold spots in V2129 Oph are both located at high latitudes
and appear at about the same rotational phases, in agreement with the
Zeeman-Doppler results \citep{don11}, but the hot spot, which produces
the veiling and some of the emission lines, is most likely located 
at the chromospheric level, more or less spatially overlapping the major
cool spot at the photospheric level.

\subsection{Emission lines}\label{emission}
\begin{figure}[htb]
\centering
\includegraphics[width=9cm]{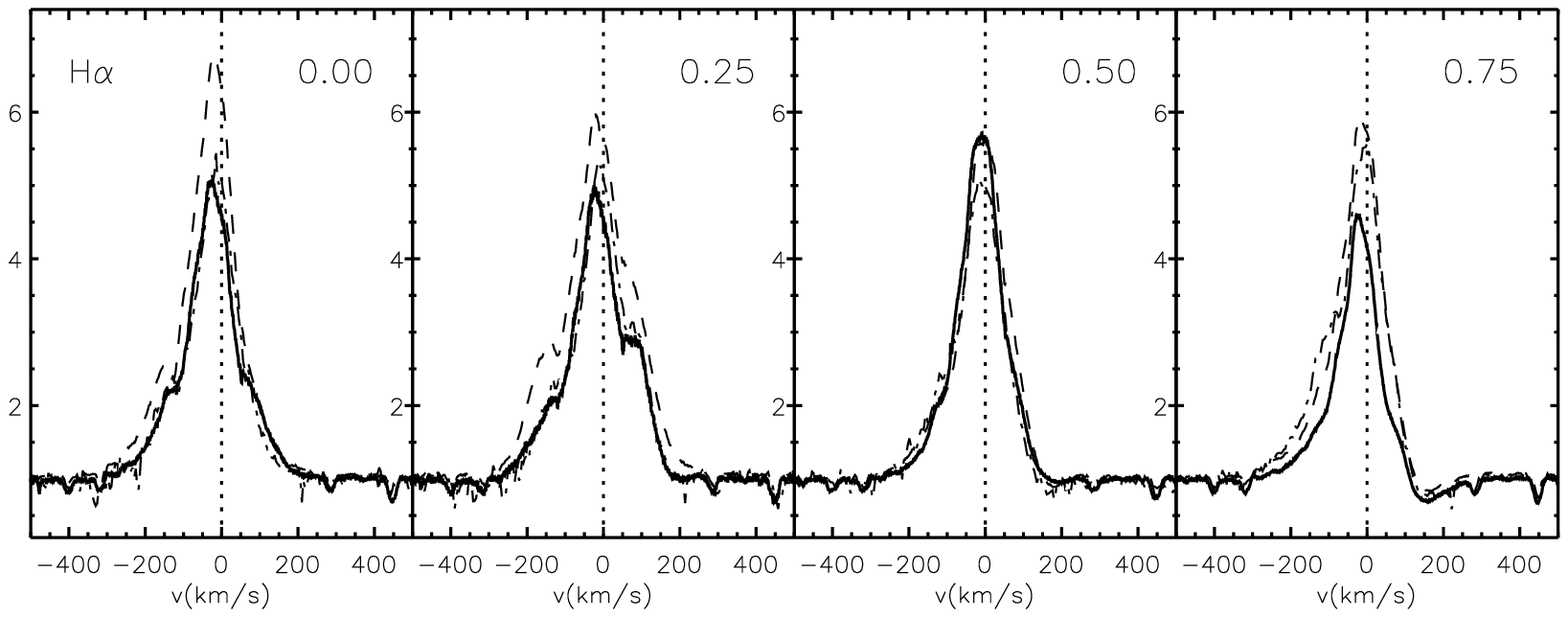}
\includegraphics[width=9cm]{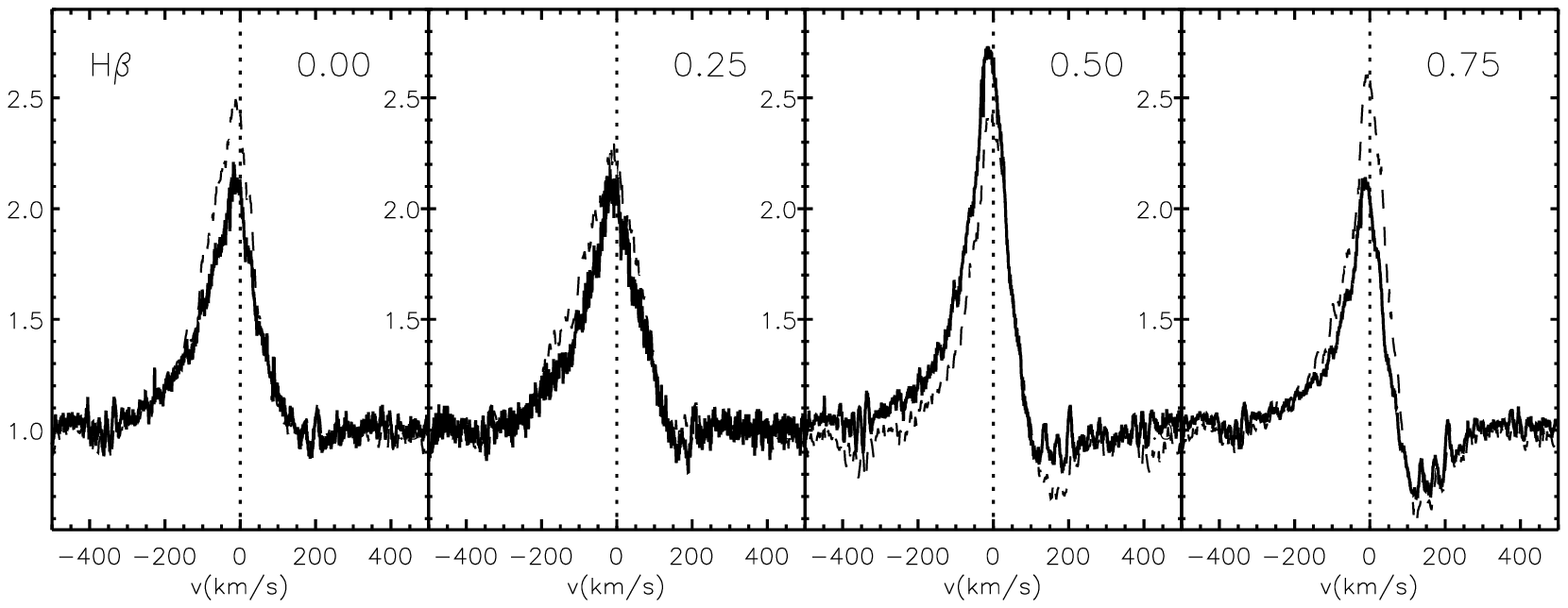}
\caption{\hal\ (top) and \hbeta\ (bottom) profiles obtained with the HARPS (solid),
ESPaDOnS (dashed), and SMARTS (dash-dotted, \hal\ only) spectrographs. The spectra correspond to a mean
profile around the phases given in the panels. The profiles are normalized
to the continuum level and the vertical dotted lines
indicate the stellar rest velocity.
}
\label{prof_halpha_selected}
\end{figure}

V2129 Oph displays Balmer and \he\ lines in emission, as usually observed 
in CTTSs (Figs. \ref{halpha_harps} and \ref{heI_harps}). 
The Balmer emission lines contain a major emission component with a
rather triangular shape that is very difficult to decompose with simple Gaussian
profiles. Blue and red emission bumps appear as the star rotates and a redshifted
absorption component is seen in both \hal\ and \hbeta, in many 
phases in the second half of the rotational cycle.
The Balmer emission-line profiles
vary substantially in phase but also exhibit a strong phase coherence,
presenting the same characteristics in similar phases, as can be seen
in Fig. \ref{prof_halpha_selected} where we show the HARPS, ESPaDOnS, and SMARTS
\hal\ and \hbeta\ emission-line profiles averaged at a few chosen phases. 

\begin{figure}[htb]
\centering
\includegraphics[width=9cm]{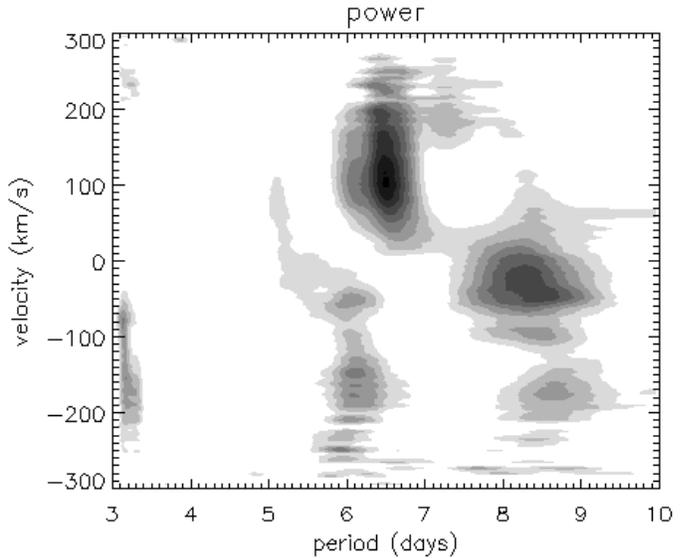}
\caption{\hal\ periodogram obtained with the HARPS, ESPaDOnS, and SMARTS
observations. The power scale ranges from 0 (white) to a maximum value
of 18.7 (black).
}
\label{per_halpha}
\end{figure}

The \hal, \hbeta, and \hei\ lines display periodic variability across the 
profiles, at periods that range typically from 6.0 to 6.9 days, which
are probably mostly driven by rotation, given the close match to the stellar 
rotation period. In Fig. \ref{per_halpha}, we show the \hal \ periodogram, where
we can see a 6.5-day-period across the red wing, while the blue wing shows 
periodicities at both 6.00 days and about 8.3 days (Fig. \ref{per_halpha}).  
We do not have an obvious interpretation of the longer period in the
blue wing of \hal \ but it could, for example, be related to a diskwind 
coming from a region outside the co-rotation radius.

The \hal\ profiles show a clear red-emission shoulder from
phases 0.05 to 0.35 and the redshifted absorption appears from
phases 0.6 to 0.8. The \hbeta\ profiles have a redshifted absorption 
component in a more extended phase range, going from 0.5 to 0.9. This
is consistent with the formation of \hbeta\ in the accretion column, closer
to the star than the source of \hal. In that way, the hot spot is seen through the 
accretion column projected along our line of sight during a more extended
phase range in \hbeta\ than in \hal, creating the redshifted absorption
component.
The redshifted absorption components are seen in both
\hal\ and \hbeta\ at phases that follow the passage of the hotspot. This indicates 
that the major accretion column must be trailing the hotspot passage. This
could happen if the stellar magnetic field were at least partially anchored to 
the disk at radii beyond the co-rotation radius.
\citet{don11} obtained a co-rotation radius of $r_{\rm cor} \simeq 7.7$ \rstar\ 
and a magnetospheric radius of 7.2 \rstar\ with the ESPaDOnS observations.
The maximum value of the redshifted absorption velocity is 
$\sim 260$ \kms\ in \hal\ and $\sim 300$ \kms\ in \hbeta. 
Although V2129 Oph was shown to have an octupolar magnetic-field component 
that is stronger than the dipolar component at the surface of the 
star, near the co-rotation radius the dipolar component is much stronger 
(by a factor of about 25) than the octupolar \citep{don11} and accretion is 
then expected to occur mostly through dipole field lines.
Assuming that the
system is seen at $i=60$\degr, the hot spot is located at $lat=80$\degr, 
and the accreting material free-falls along a dipolar accretion column,
we obtain a truncation radius of $7.8$ \rstar. 
This value is consistent with the possibility that part of 
the stellar magnetic field may interact with the disk beyond the co-rotation radius,
creating a trailing accretion column in the 2009 observations.

The \hal\ and \hbeta\ lines showed an increase in both intensity and width one 
day before (JD=4998) the \hei\ line, the mass accretion rate, and the veiling all increased 
significantly (Figs. \ref{heI_harps}, \ref{macc}, \ref{veiling_heIvrad}). 
If \hal\ and \hbeta\ are produced in a more extended region than \hei,
they could be affected by an increase in the mass accretion rate before the hot spot
(veiling) and the \hei\ line, which trace regions close to the accretion shock.
Interestingly, the day \hal\ and \hbeta\ increased, the \hei\ line, which normally 
has only a slightly redshifted ($v_{\rm peak} \sim 4$ \kms) narrow (FWHM $\sim 20$ \kms)
component in V2129 Oph, showed a small broad component (BC), as can be seen in
the two panels with phases 0.36 and JD=4998 of Fig. \ref{heI_harps}. We decomposed
these \hei\ profiles using two Gaussians, and obtained BCs centered at $\sim -20$ \kms \
and $\sim -17$ \kms \ that have FWHM of $71$ \kms\ and $69$ \kms, respectively.
The BC is generally associated with hot winds \citep{ber01,edw03}, hence this might 
indicate that an accretion
burst has affected both the diskwind and the Balmer lines about one day before reaching the
stellar surface. There was another burst episode in our data at JD=4993, where the
\hei \ lines are again very strong. Unfortunately, we do not have, however, spectra taken
one day before, to verify whether \hal \ and \hbeta\ had also increased in advance of \hei.

We have discussed the many characteristics of the Balmer emission lines, and we now
attempt to reproduce them with theoretical line profiles, based on a 3D MHD circumstellar structure 
calculated for V2129 Oph.

\subsection{Theoretical line profiles}\label{models}
Theoretical emission-line profiles of CTTSs have often been computed in the 
literature and compared to the observations to determine the main 
characteristics of the circumstellar accretion flow of these objects
\citep{muz98,muz01,kur06}. However, the steady-state axisymmetric dipolar models 
that are more commonly used to describe the stellar magnetosphere 
cannot reproduce the variability observed in the emission profiles of CTTSs. 
These are mainly driven by the rotational modulation of non-axisymmetric multipolar 
fields and non-steady accretion processes, which are not straightforward to model. 
Progress has been made in the determination of the surface magnetic 
fields of CTTSs \citep{don07,hus09,gre11} 
and the development of 3D MHD simulations that include non-axisymmetric 
multipolar components of the stellar magnetic field \citep{lon07,lon08,rom08}. It is now 
possible to compute emission line profiles from the funnel flow structure determined by
3D MHD simulations that are based on observed magnetic field configurations of CTTSs 
\citep{kur08,kur11}. 

The observed \hal\ and \hbeta\ profiles of V2129 Oph cannot be easily decomposed
with the simple Gaussian components typically used to represent the main accretion flow,
the wind, and the accretion shock, in a similar way to the analyses of other stars, such as AA Tau
\citep{bou07}.
To understand the observed emission-line variabilities, we therefore decided to 
try to reproduce the Balmer line profiles with radiative transfer calculations based on 
the 3D MHD structure of the accretion flow.
The \hal\ and \hbeta\ profiles were calculated with the stellar parameters 
of V2129 Oph and two different magnetic field configurations: a pure dipolar 
field and the dipole+octupole configuration proposed by \citet{don11} 
for the 2009 observations. 

The 3D MHD code and model used here were developed and described earlier
in \citet{kol02} and \citet{rom03, rom04, rom08}. 
The code has been modified by \citet{lon07,lon08} to 
incorporate the quadrupole and octupole components, and 3D modeling of accretion onto 
BP Tau and V2129 Oph, with close to realistic fields, has been successfully performed 
\citep{lon11,rom11}. Here, we computed a special set 
of simulations for V2129 Oph, where the stellar magnetic field was approximated with a 
superposition of dipole and octupole fields, and the field values 
correspond to the more recent spectropolarimetric observations of this star 
\citep{don11}.
The parameter values used in the simulations are listed in 
Table \ref{tab_models}. The inclinations of the dipole and the octupole
with respect to the rotation axis were 15\degr \ and 25\degr, respectively, 
and the octupole is about 0.1, in rotational phase, ahead of the dipole,
as proposed by \citet{don11}. 
The mass accretion rate used in the model, $6.3 \times 10^{-10}$ \msun yr$^{-1}$, 
is slightly lower than the lowest values measured for the system 
($9 \times 10^{-10}$ \msun yr$^{-1}$). This low accretion rate was necessary
to generate velocity profiles that were compatible with the observed ones,
since increasing the mass accretion rate would bring the disk closer to the star
and produce narrower emission-line profiles.
The simulations were done in dimentionless form, and the 
current simulations are applicable if we increase the dipole component
of the field and accordingly increase the mass accretion rate. For
example, the MHD simulation run would correspond to the observed mass accretion 
rate, if the dipole component of the field were increased by a factor of 1.2,
such that the dipole field were 1.08 kG instead of 0.9 kG.

\begin{figure}[htb]
\centering
\includegraphics[width=9cm]{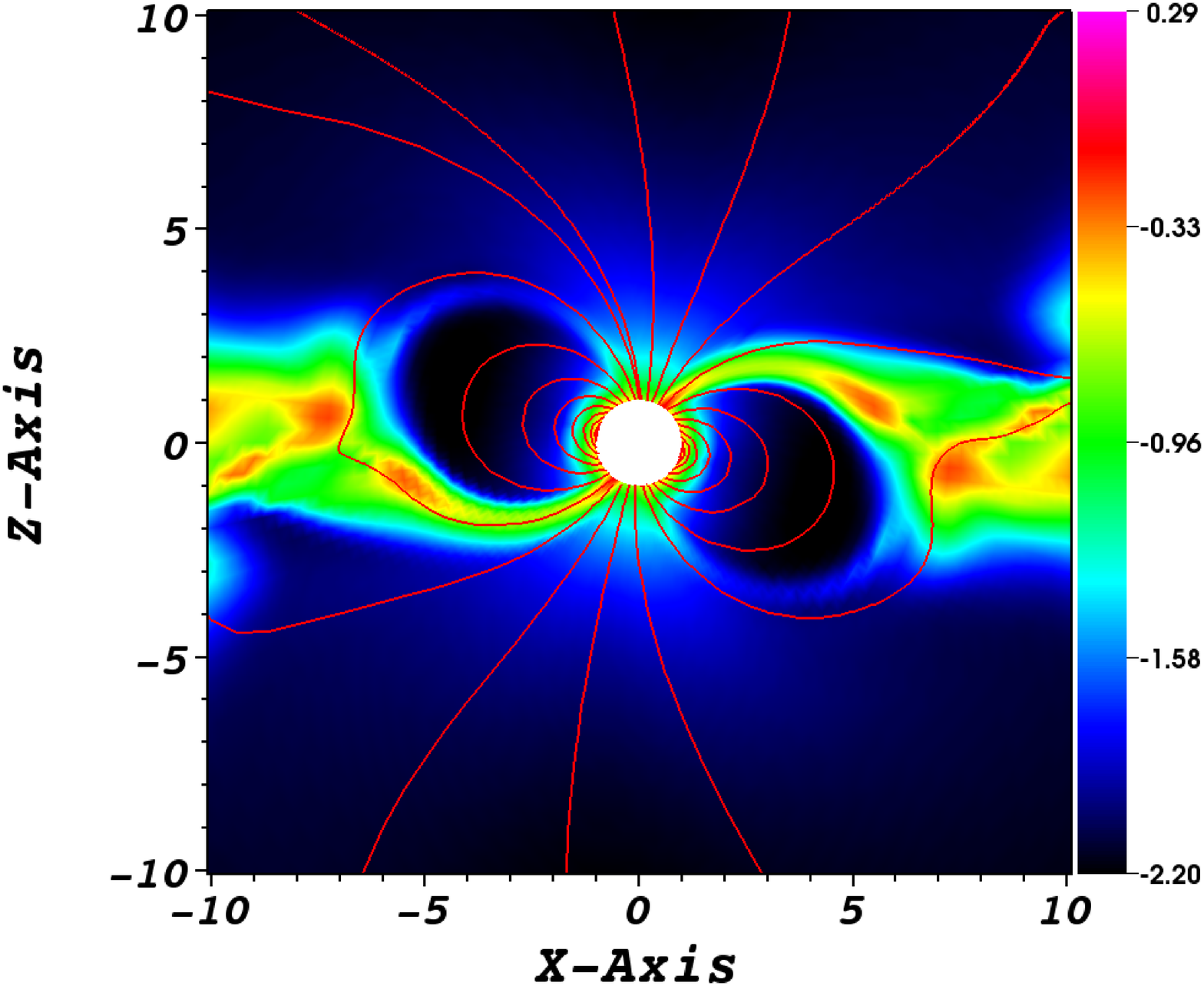}
\includegraphics[width=9cm]{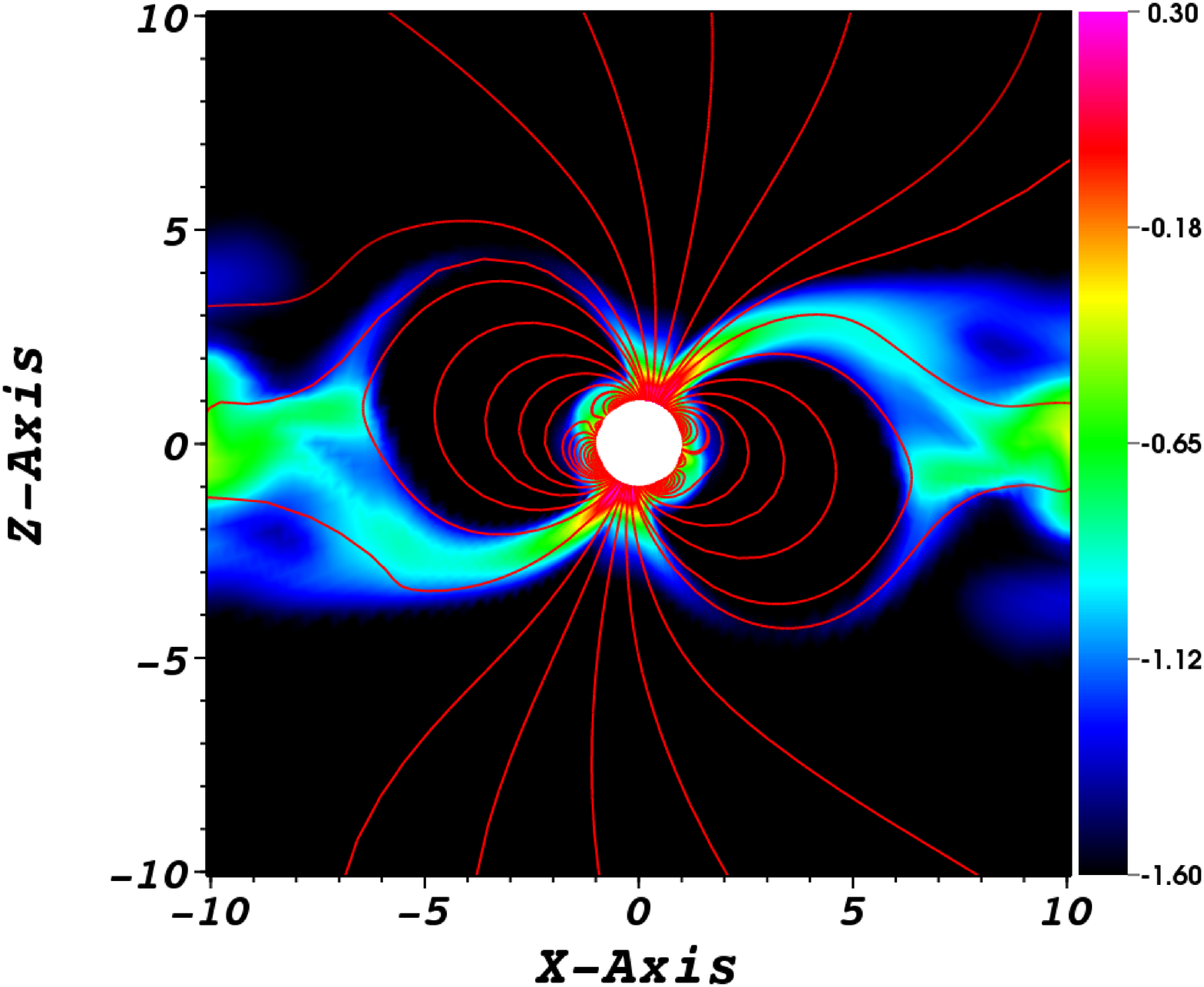}
\caption{Density slice in the $xz$ plane for the 3D MHD simulations with
dipole-only (top) and dipole+octupole (bottom) magnetic fields. 
The parameters used in the simulations are 
listed in Table \ref{tab_models}.
The length scales are in units of the stellar radius ($R_{*}$) and 
the intensity is shown on a logarithmic scale with an arbitrary unit.}
\label{density_slice}
\end{figure}

The three-dimensional simulations show that, in a pure dipole case, matter flows in two ordered funnel 
streams towards the closest magnetic pole, as can be seen in 
Fig. \ref{density_slice} (top panel), which shows a slice of the density distribution 
in the $xz$ plane and selected magnetic field lines. The disk is truncated by 
the magnetosphere at radius $r_t\approx 6-7$ \rstar, and matter 
hits the star below the magnetic pole to form two bean-shaped 
hot spots. In the dipole+octupole case (Fig. \ref{density_slice}, bottom panel), 
the dipole component strongly 
dominates the field at large distances from the star, determining the 
truncation of the disk at similar distances as in a pure dipole case. However, 
close to the star, the octupole component dominates the flow and the octupolar field 
redirects the funnel to higher latitudes, where matter hits the star closer to the octupolar 
magnetic pole \citep[see also ][]{rom11}. This has an effect in the hot spot location. In a pure 
dipole case, the hot spot forms at some distance (about 20\degr) from the magnetic pole. 
Hence, the hot spot and the cold spot, which is associated with the magnetic pole, should 
not coincide in phase. However, in the dipole+octupole case the hot spot is located 
much closer to the magnetic pole, because the octupolar component redirects the 
initial funnel flow to higher latitudes, which agrees with the observational results
presented in Sect. \ref{spots} and in \citet{don11}. 

Next, we calculated spectral lines from the MHD modeled flows. We followed 
the same approach described in \citet{kur08} to obtain emission line profiles 
from the density, velocity, and temperature structures of the MHD simulations. 
These distributions were mapped on to the radiative transfer grid of the TORUS code 
\citep{har00,kur06,kur11}, and the corresponding line source functions were 
calculated. The theoretical profiles of \hal \ and \hbeta \ were computed
using the Sobolev escape probability method, as described in 
\citet{kle78} and \citet{har94}.

\begin{table}
\caption{Parameters used in the MHD simulations}\label{tab_models}
\begin{tabular}{llllllll}
\hline \hline
\mstar & \rstar & i & $\log{{\dot{M}}}$ & $R_{\rm c}$ & $P$ & $B_{\rm dip}$ & $B_{\rm oct}$\\
(\msun) & (\rsun) & (deg) & (\msun ${\rm yr}^{-1}$) & (\rstar) & (days) & (kG) & (kG)\\
\hline
1.35  & 2.1 & 60  & $-9.2$  & 7.7 & 6.53 & 0.9 & 2.1\\
\hline
\end{tabular}
\end{table}

\begin{figure}[htb]
\centering
\includegraphics[width=9cm]{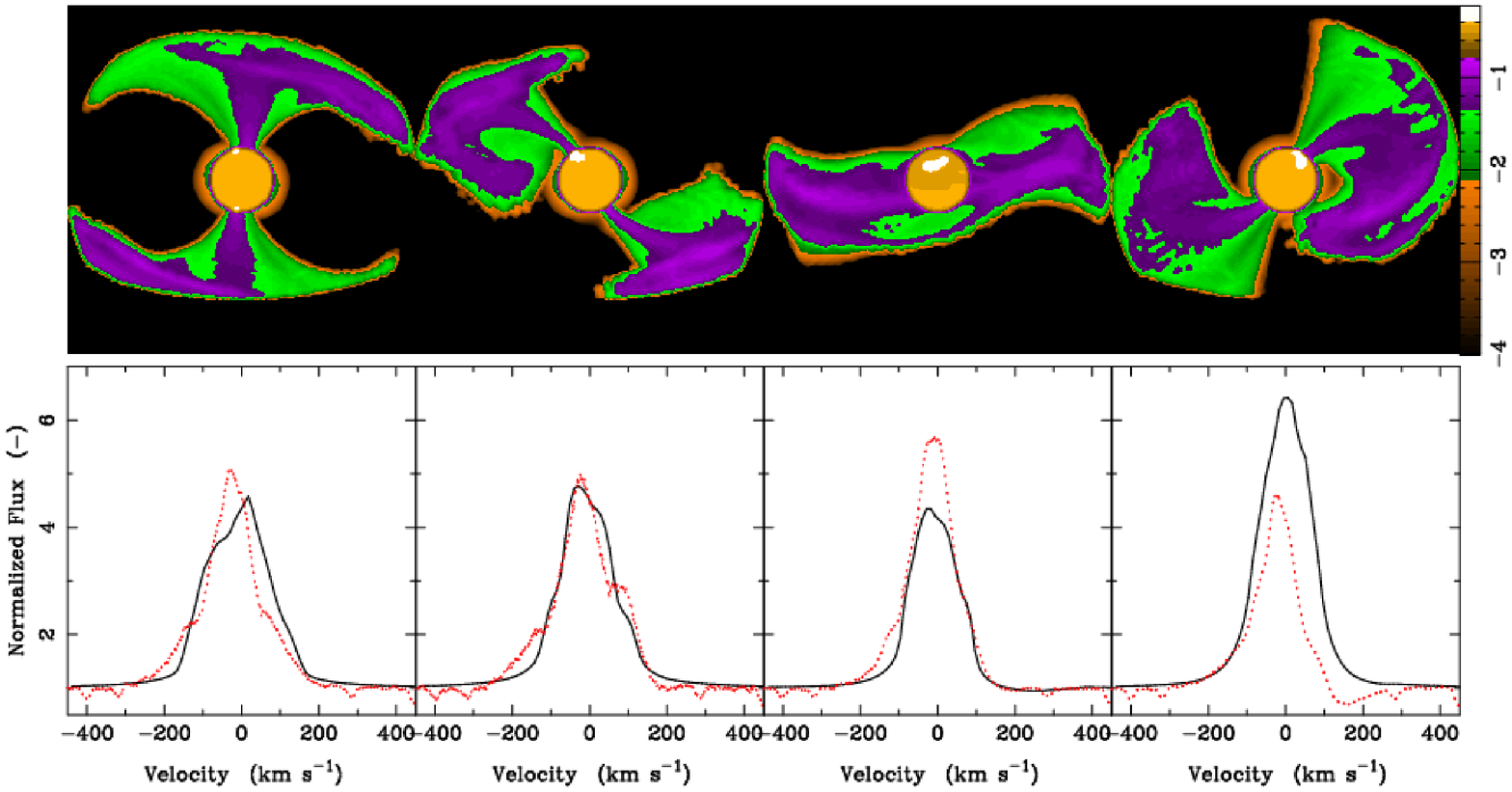}
\caption{\hal \ model intensity maps (top panels) and the corresponding profiles 
(bottom panels, solid lines) computed at rotational phases of 0.0, 0.25, 0.5, and 0.75 
(from left to right) for the 3D MHD simulation with a dipole-only magnetic field
(see Fig. \ref{density_slice}). The intensity maps are shown in the plane perpendicular
to our line-of-sight. The parameters used in the simulations are listed in Table \ref{tab_models}. 
The intensity is shown on a logarithmic scale with an arbitrary unit.
The model profiles (solid lines) are compared with the observed mean \hal \ profiles
(dotted lines) obtained with HARPS that are presented in Fig. \ref{prof_halpha_selected}.
The profiles are normalized to the continuum level.
}
\label{prof_halpha_model_dipole}
\end{figure}

\begin{figure}[htb]
\centering
\includegraphics[width=9cm]{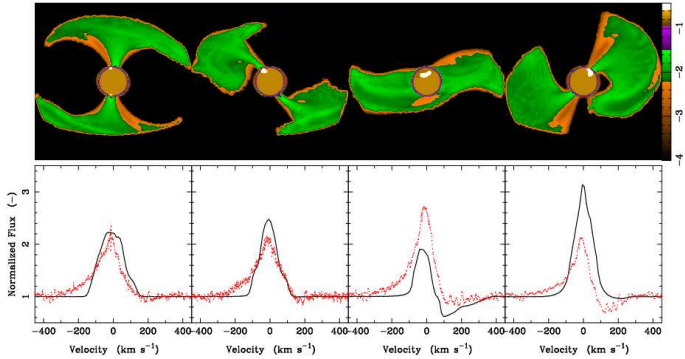}
\caption{Same as in Fig. \ref{prof_halpha_model_dipole}, but for \hbeta.
}
\label{prof_hbeta_model_dipole}
\end{figure}

\begin{figure}[htb]
\centering
\includegraphics[width=9cm]{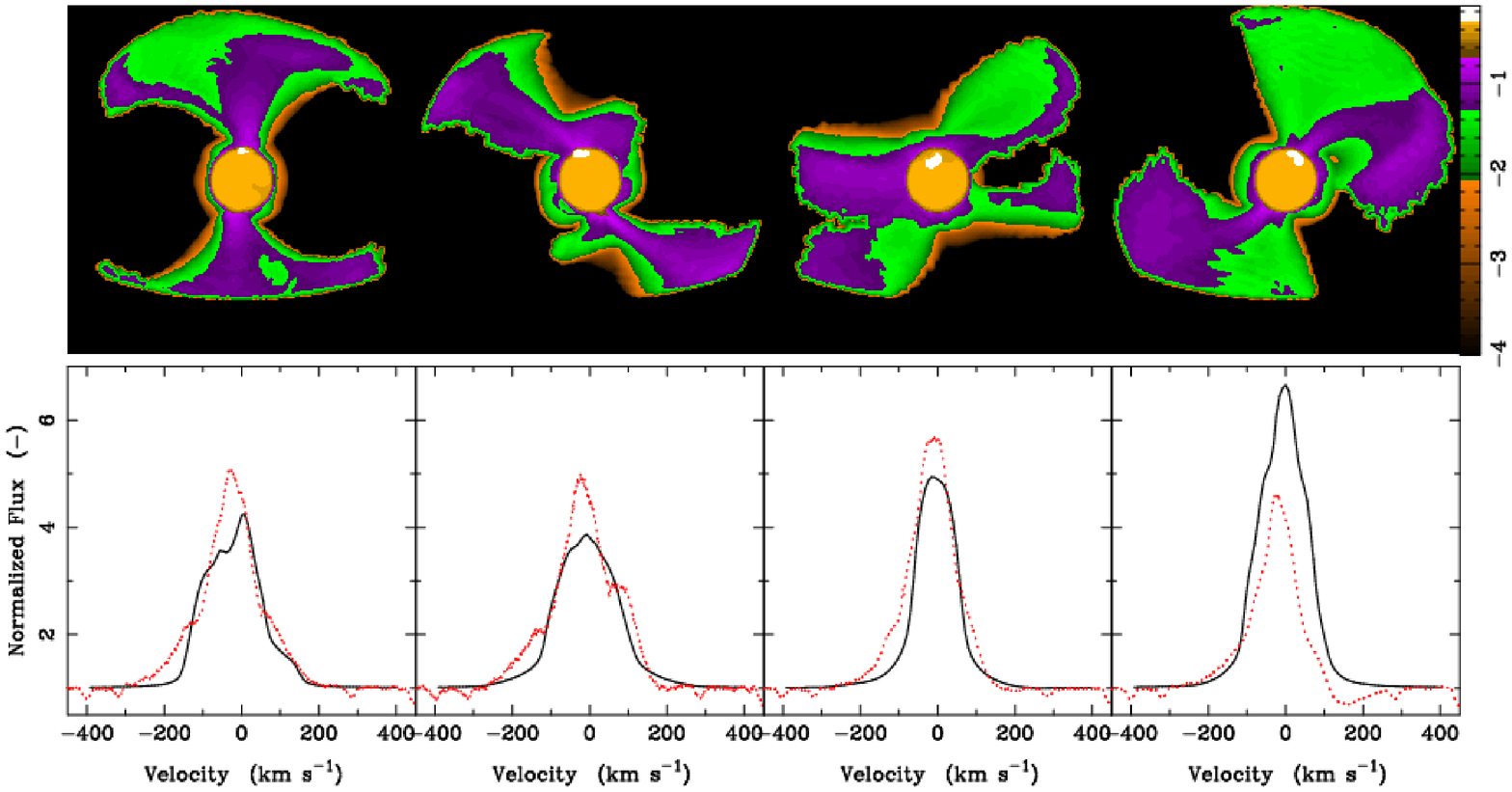}
\caption{\hal \ model intensity maps (top panels) and the corresponding profiles
(bottom panels, solid lines) computed at rotational phases 0.0, 0.25, 0.5, and 0.75
(from left to right) for the 3D MHD simulation with a dipole+octupole magnetic field
(see Fig. \ref{density_slice}). The intensity maps are shown in the plane
perpendicular to our line-of-sight. The parameters used in the simulations are listed
in Table \ref{tab_models}. The intensity is shown on a logarithmic scale with an
arbitrary unit. The model profiles (solid lines) are compared with the observed
mean \hal \ profiles (dotted lines) obtained with HARPS and
shown in Fig. \ref{prof_halpha_selected}.
The profiles are normalized to the continuum level.
}
\label{prof_halpha_model}
\end{figure}

\begin{figure}[htb]
\centering
\includegraphics[width=9cm]{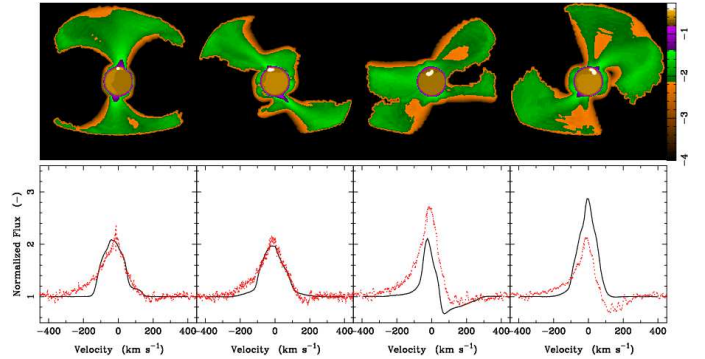}
\caption{Same as in Fig. \ref{prof_halpha_model}, but for \hbeta.
}
\label{prof_hbeta_model}
\end{figure}

Our initial 3D MHD calculations were done with an adiabatic index of $\gamma=5/3$
and, although the \hal \ and \hbeta \ line profiles computed with the dipole-only 
magnetosphere were quite similar to the observations,
the profiles obtained with the dipole+octupole magnetic field configuration 
were much narrower than the observed ones.
The 3D simulations do not include radiative cooling, but only 
adiabatic cooling and heating. This leads to funnel temperatures that
are too high, relative to observational values, and generates high
pressure and pressure gradients in the funnel, which slow down the accretion. 
These effects were more pronounced in the presence of the octupole component, which,
with its many loops close to the stellar surface (Fig. \ref{density_slice}, 
bottom panel), acted as a wall against accretion.
We then adjusted the adiabatic index of the gas in the equation of state to 
a lower value ($\gamma=1.2$ instead or 5/3) to mimic a cooling effect. 
This helped to decrease the thermal pressure of the gas near the surface, 
and to produce a faster flow near the star. 
With the above adjustments, the flow geometry, on a large scale, for the
dipole+octupole case (Figs. \ref{prof_halpha_model} and \ref{prof_hbeta_model},
top panels) is somewhat similar to that of the dipole-only case
(Figs. \ref{prof_halpha_model_dipole} and \ref{prof_hbeta_model_dipole}, top panels).

\begin{figure}[htb]
\centering
\includegraphics[width=9cm]{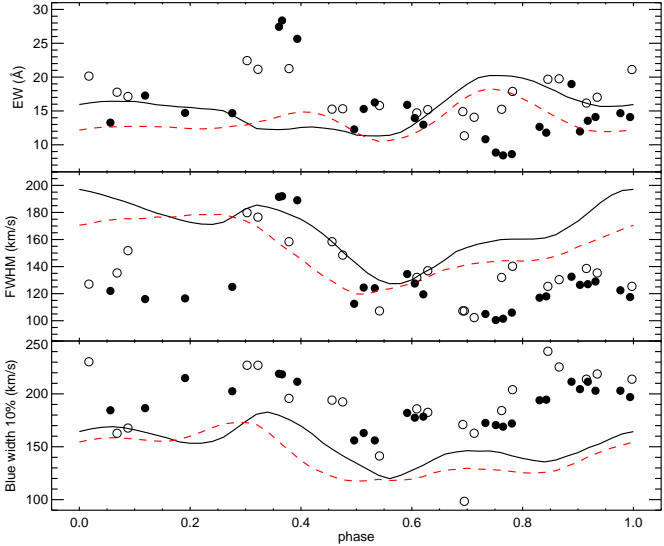}
\caption{\hal \ line EW (top), full width at half maxima (middle),
and blue side width at 10\% of the peak intensity (bottom) as a function
of rotational phase. HARPS and ESPaDONs data are shown as filled and open
dots, respectively. The values measured in the profiles of the dipole-only 
(solid line) and dipole+octupole (dashed line) models are overplotted. 
}
\label{prof_halpha_characteristics}
\end{figure}

\begin{figure}[htb]
\centering
\includegraphics[width=9cm]{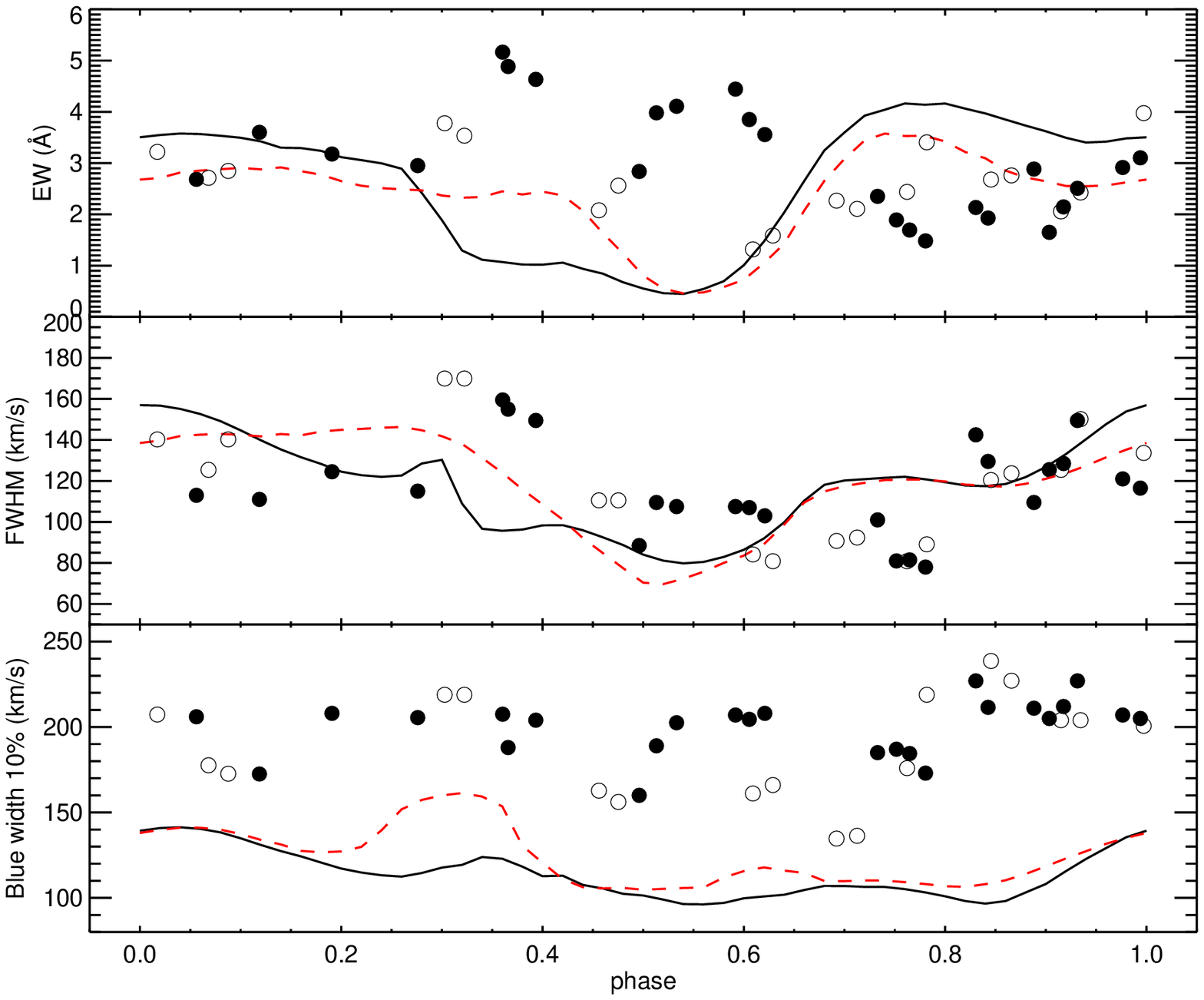}
\caption{Same as in Fig. \ref{prof_halpha_characteristics}, but for \hbeta.
}
\label{prof_hbeta_characteristics}
\end{figure}

In the bottom panels of Figs. \ref{prof_halpha_model_dipole}, 
\ref{prof_hbeta_model_dipole},   \ref{prof_halpha_model}, and \ref{prof_hbeta_model}, 
we show the mean \hal \ and \hbeta\ observed profiles at selected phases, overplotted on the mean 
theoretical line profiles at the same phases.
We can see that the Balmer line profiles obtained with the models 
resemble overall the observed
profiles for both \hal\ and \hbeta\ and for both the dipole-only and
dipole+octupole models. 
In Figs. \ref{prof_halpha_characteristics}
and \ref{prof_hbeta_characteristics}, we show the theoretical and observed line EWs,
the full width at half maxima (FWHM), and the width on the blue side at 10\% peak
intensity for both \hal\ and \hbeta. We only use the blue side width at 10\% peak
intensity in our comparison, since the red side width at 10\% peak intensity is 
strongly affected by the redshifted absorption component that is not very accurately 
modeled (see discussion later in this section).

In Figs. \ref{prof_halpha_characteristics} and \ref{prof_hbeta_characteristics} 
(top panels), we see that the \hal\ and \hbeta\ theoretical EWs are of 
the same order of magnitude as the observed ones. However, the variability predicted 
by the models for the EW of both lines is quite different from the observations.
On the other hand, the theoretical \hbeta\ FWHM is very similar to the observed values, 
both qualitatively and quantitatively, while the theoretical FWHM of \hal\ are a bit overestimated 
when compared to the observations.
The \hal\ and \hbeta\ theoretical emission-line widths vary during the rotational cycle, being 
wider at phases 0.0 and 0.25 than at phases 0.5 and 0.75, as observed.
However, as seen in the bottom panels of Figs. \ref{prof_halpha_characteristics} and 
\ref{prof_hbeta_characteristics}, although the qualitative behavior of the line width is similar
between models and observations, the model profiles are much narrower than the observed ones.
This is also clearly seen in the bottom panels of
Figs. \ref{prof_halpha_model_dipole}, \ref{prof_hbeta_model_dipole},
\ref{prof_halpha_model}, and \ref{prof_hbeta_model}.
In these figures, we can see that the intensity and shape of the theoretical and observed 
emission lines closely resemble each other, especially around phases 0.0 and 0.25. 
At phases 0.5 and 0.75, the match is not so good, the line strengths
of the models being slightly weaker and stronger than those of the
observations at phases 0.5 and 0.75, respectively. The differences in the line 
strengths at phases 0.5 and 0.75 may be attributed to differences between 
the flow geometry of the MHD models and the real system.
The extension of the red wing is similar in models and observations
and both the dipole-only and dipole+octupole models display clear
redshifted absorption going below the continuum and extending up to
300-350 \kms \ in \hbeta\ at phase 0.5, in agreement with the observations.
In the dipole-only models, we also see some hint of redshifted 
absorption in \hbeta\ at phase 0.75, as observed,
while no absorption is seen at these phases in the dipole+octupole models.
Most of the line variability and the appearance of  
redshifted absorption in the Balmer lines can therefore be attributed to changes in the observer's 
viewing angle of the system due to a combination of the rotational motion 
of the star and its inclined and non-aligned magnetosphere.

A noticeable difference between models and observations is
the strength and extension of the blue wings of the Balmer line profiles 
(especially in \hbeta), as can be seen in the bottom panels of 
Figs. \ref{prof_halpha_model_dipole}, \ref{prof_hbeta_model_dipole}, 
\ref{prof_halpha_model}, and \ref{prof_hbeta_model}.
The emission in the blue wings extends to $\sim 300\,\mathrm{km\,s^{-1}}$ in 
the observations, but it extends only to $\sim 150\,\mathrm{km\,s^{-1}}$ in most of 
the \hbeta\ theoretical profiles. An exception is the dipole+octupole model at a 
phase of around 0.25, which shows a weak extension of the blue wing emission up to $\sim
300\,\mathrm{km\,s^{-1}}$. 
Possible causes of this discrepancy are (1) the uncertainty in the flow geometry, 
(2) the inclination angle, or (3) the presence of non-negligible wind emission.  
Although the observations do not show any clear sign of an outflow/wind, such 
as a blueshifted absorption component, the wind may contribute to the emission. Unfortunately, 
the wind is not included in our current MHD models.

Another difference is that the observations show clear redshifted absorption
in both \hal\ and \hbeta, while in the present model, there is clear 
redshifted absorption only in \hbeta.
The redshifted absorption in \hbeta\ also appears at lower velocities in 
the simulations, starting at 50 \kms, than in the observations, 
starting at 100 \kms, which reflects the inflow speed of the gas in the 
adopted MHD solution. 
Another small difference between the models and observations is that, in the MHD simulation, 
the major funnel flow is centered around
the hot spot, instead of mostly trailing it, as seen in the observations. In the dipole-only simulations,
the \hbeta\ redshifted absorption appears from phases 0.40 to 0.68. In the 
dipole+octupole simulations, it appears from phases 0.44 to 0.64, almost centered
at phase 0.5, when the hot spot faces the observer, while in the observations it appears
from phases 0.45 to 0.9 (see Sect. \ref{emission}). 
This difference is probably caused by the magnetosphere being inside the 
co-rotation radius ($R_c=7.7$ \rstar), going from about 5.5 \rstar \ 
to 7.5 \rstar \ in the 3D MHD simulations (Fig. \ref{density_slice}, top). The inner radius in the
simulations was estimated from the location of the edge of the funnel flow, and the outer 
radius from the location of the largest closed magnetic field line. The outer magnetosphere radius 
is actually quite close to the magnetospheric radius derived by \citet{don11} (7.2 \rstar).
However, since it is smaller than the co-rotation radius, it is inconsistent with a trailing funnel flow
and it might instead represent a funnel flow that leads rather than trails the hot spot in the simulations.
A way to increase the phase range where redshifted absorption appears in the theoretical profiles
would be to either decrease the tilt of the dipole component or increase the mass accretion rate of the
system. Our MHD simulations usually show that, decreasing the tilt of the dipole component,
the magnetic funnel becomes wider in the azimuthal direction and the width of the funnel curtain 
also depends on the density level \citep{rom03}. At lower density levels, the magnetic curtain
covers a wider range of azimuthal angles, even at the present dipole tilt. Hence,
at a somewhat higher mass accretion rate, we may have a wider curtain at the density 
level required to obtain the redshifted absorption in \hbeta. This would make the
redshifted absorption appear in a more extended phase range.

\begin{figure}[htb]
\centering
\includegraphics[width=4.4cm]{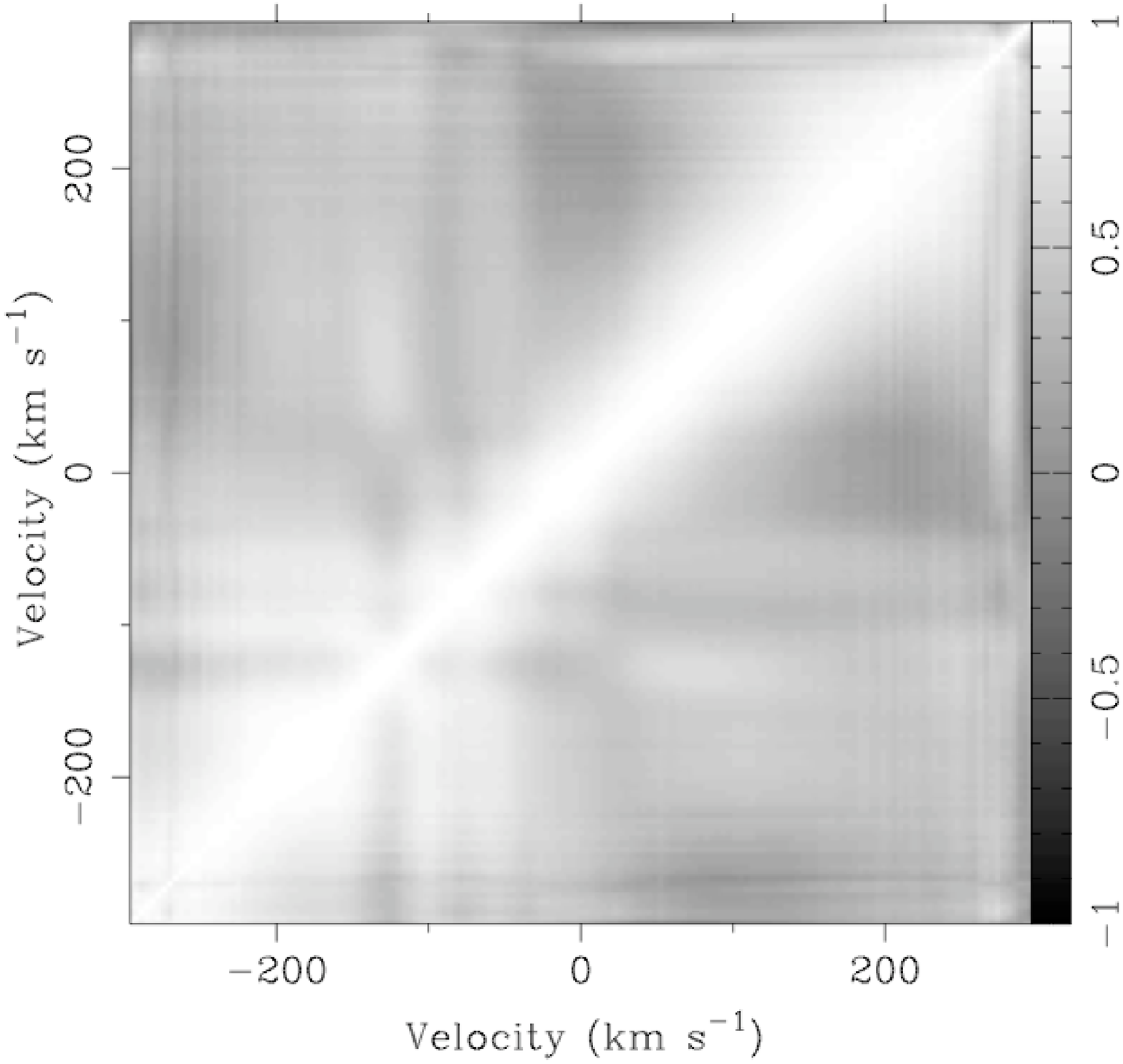}\includegraphics[width=4.4cm]{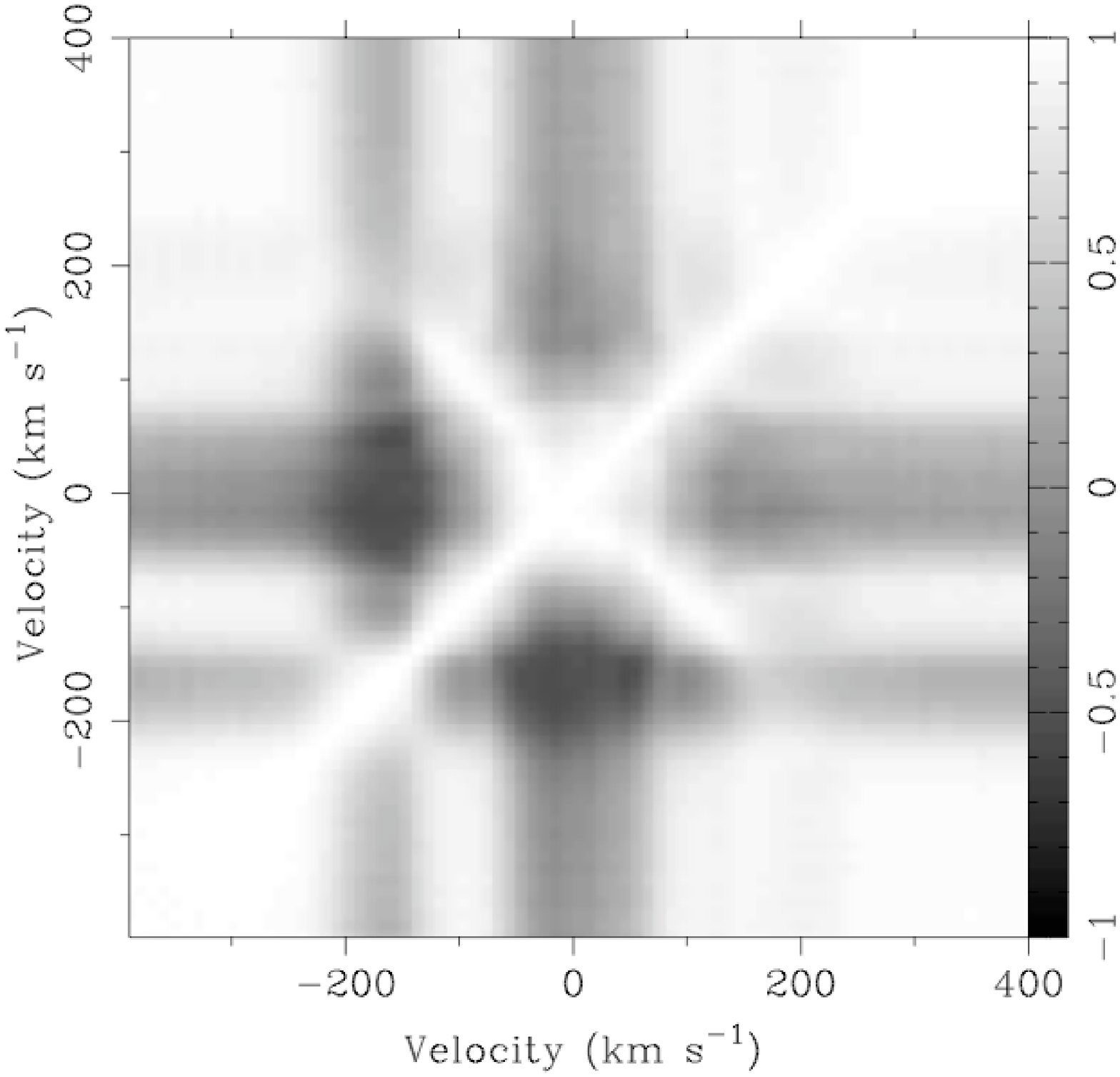}
\includegraphics[width=4.4cm]{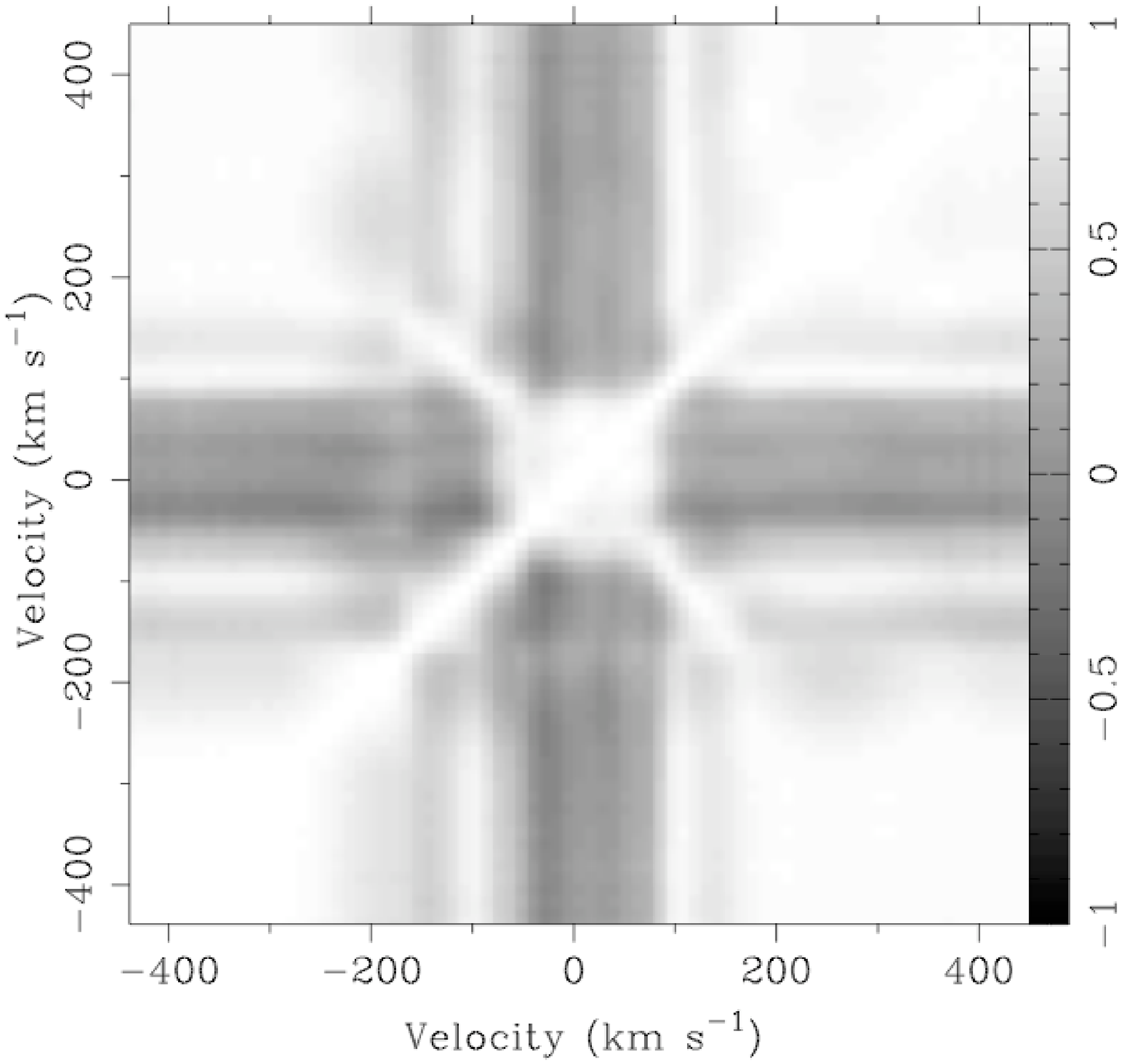}\includegraphics[width=4.4cm]{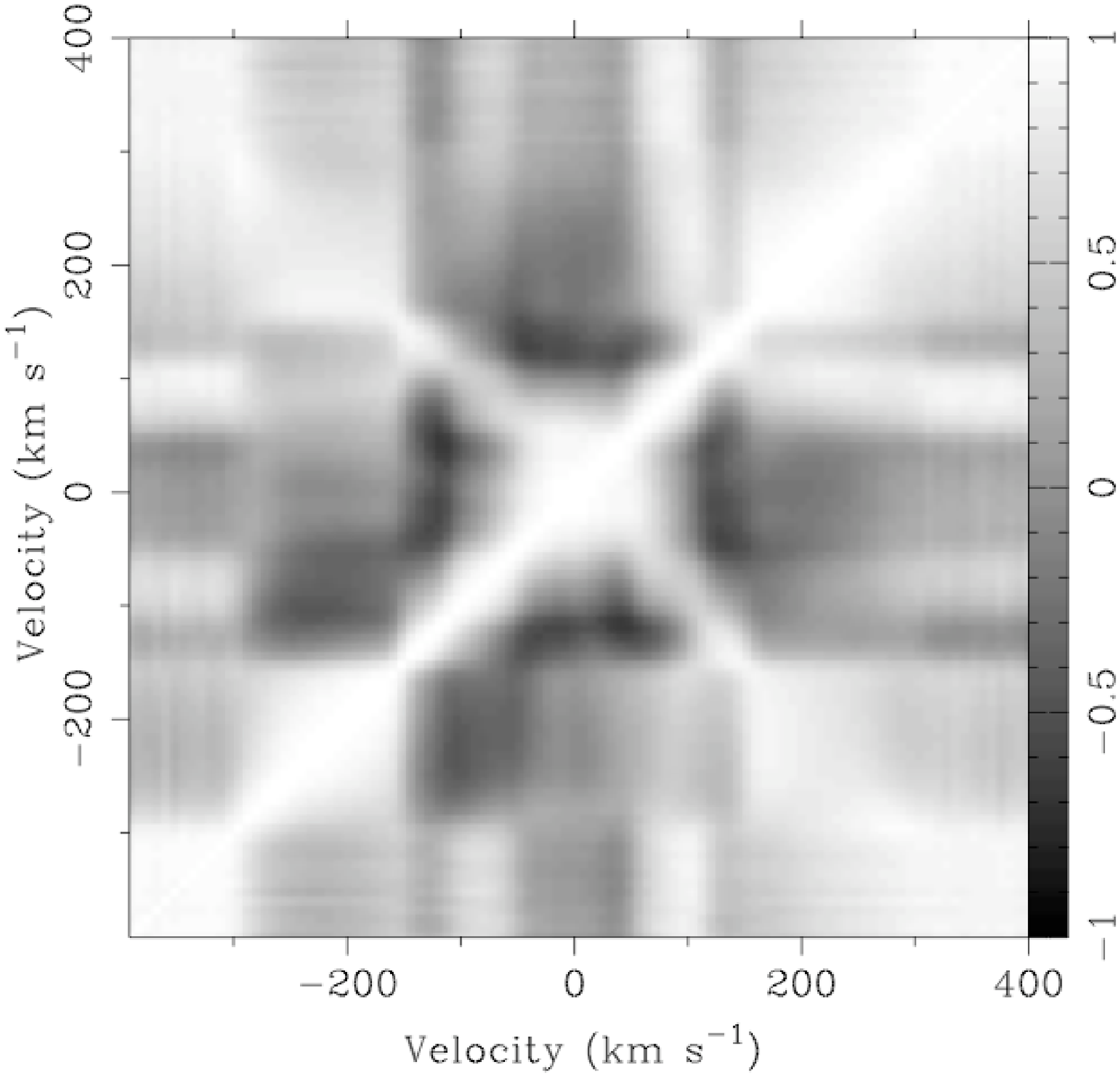}
\caption{\hal \ autocorrelation matrices. The gray scale represents
the linear correlation coefficient ($r$). White corresponds to a positive correlation
($r=1$) and black to an anti-correlation ($r=-1$). {\it Top left}: observed data.
{\it Top right}: dipole-only model with $\gamma=5/3$.
{\it Bottom left}: dipole-only model with $\gamma=1.2$.
{\it Bottom right}: dipole+octupole model with $\gamma=1.2$.
}
\label{matrices}
\end{figure}

The observed line profiles may come from different emitting/absorbing regions, which
is specially true for \hal, and it is interesting to see how these different regions
are correlated.
To evaluate the correlation variability across the line profile, we computed
autocorrelation matrices for \hal \ with the observed and theoretical line profiles.
The results are presented in Fig. \ref{matrices}.
\citet{don07} found an anti-correlation between the red wing ($v > 100$ \kms) and
the central emission peak in their ESPaDOnS data taken over seven days, covering one 
rotational cycle. They attributed the anti-correlation to the contributions at different 
velocities to the \hal \ profile from distinct physical regions (accretion pre-shock, 
winds, shock region). 
With the present data, we are unable to see such an anti-correlation between the red wing and 
the emission peak, and the two regions actually do not seem to be strongly correlated,
which agrees with the matrices calculated with the theoretical profiles. 
This could indeed be due to the redshifted absorption component coming from the pre-shock region 
of the accretion funnel, while the central part of the emission profile might have contributions
from either a much larger volume of the accretion column or a wind. They could then 
represent independent variations. The differences between the two observational sets (\citet{don07}
and ours) could be due to the larger number of rotational cycles covered with the present data set 
that would tend to smooth out any peculiarities in a single rotational cycle.
The observed \hal \ autocorrelation matrix bears some resemblance to the dipole-only 
autocorrelation matrix (especially the $\gamma=5/3$ model). Although the degree of correlation is not
exactly the same across the line in both matrices, most of the correlated/anti-correlated
regions appear in both the dipole-only and the observed matrices. Alternatively, the
dipole+octupole matrix contains many correlated/anti-correlated regions that
are not present in the matrix obtained with the observed data.

\begin{figure}[htb]
\centering
\includegraphics[width=4.3cm]{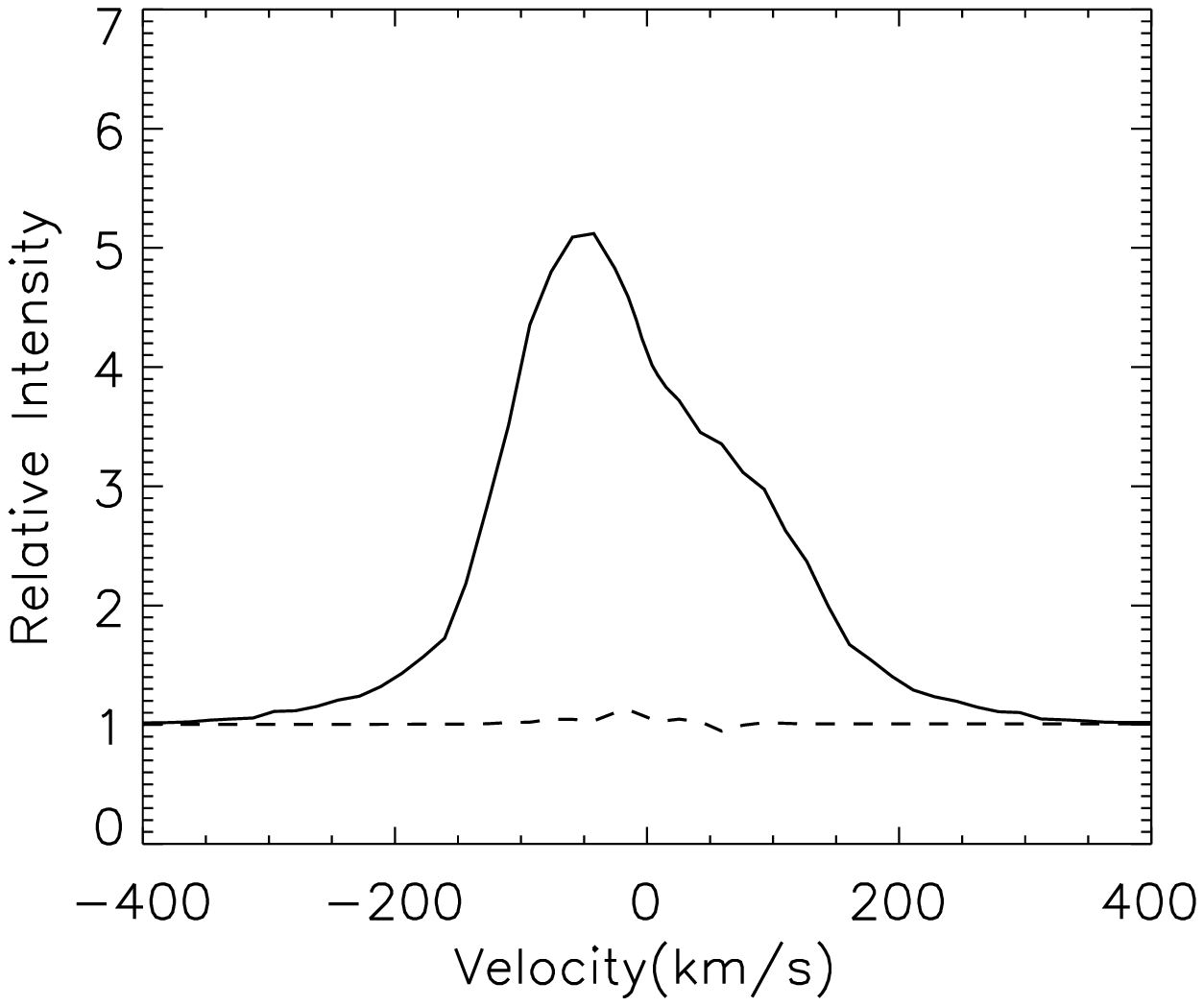}\includegraphics[width=4.3cm]{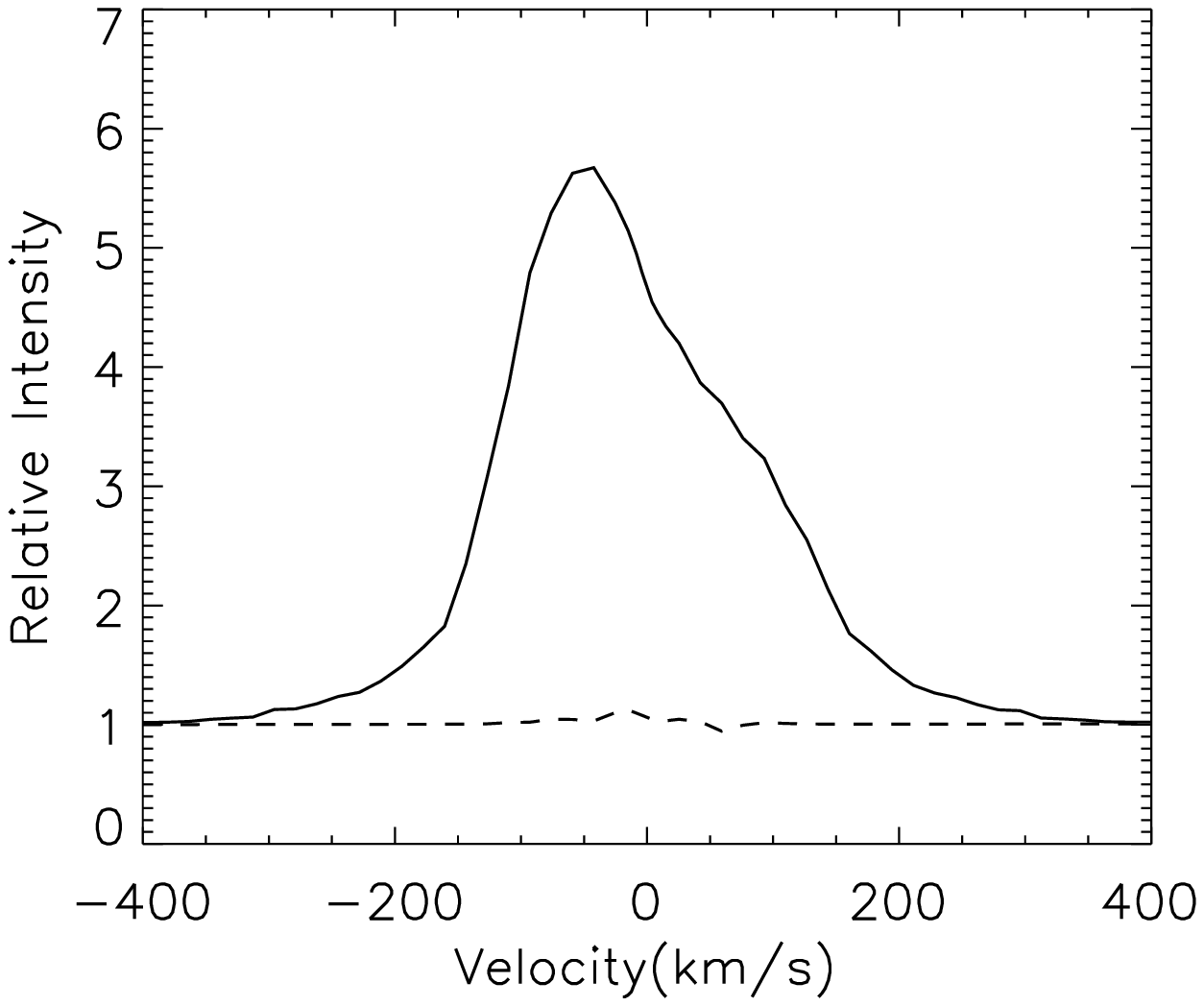}
\caption{\hal \ profiles calculated with the dipolar and axisymmetric hybrid model that includes both magnetospheric
accretion and diskwind (solid lines). The dashed lines show only the diskwind contribution to each profile. 
We used the parameters presented in Table \ref{tab_models}, 
a maximum wind temperature of $T_{\rm wind,max}=9000$ K and a photosphetric temperatures of 
$T_{\rm phot}=4500$ K. We varied the maximum temperature of the magnetosphere from 
$T_{\rm mag,max}=9000$ K (left) to $T_{\rm mag,max}=8950$ K (right).
}
\label{wind}
\end{figure}

None of our theoretical models have included a wind component, which is expected
to be present in these systems, even if it does not always contribute significantly
to the line profile \citep{kur06,lim10}.
There are some blueshifted features, such as a shoulder, in the observed \hal \ profiles that could be 
due either to an absorption, coming from a wind, or to an emission excess, from different 
projections of the accretion funnel. 
To verify the importance of a diskwind to the observed \hal \ line,
we calculated theoretical \hal \ profiles with the hybrid MHD model of \citet{lim10},
which includes both magnetospheric and diskwind components. In this model, the magnetosphere
is an axisymmetric dipole and we aim to reproduce the average observed profile of V2129 Oph,
using the parameters presented in Table \ref{tab_models}. The magnetosphere extends from
7.2 \rstar \ to 7.8 \rstar, where we kept the maximum temperature of the magnetosphere around 
$T_{\rm mag,max}=9000$ K and used the photospheric temperature of $T_{\rm phot}=4500$ K. Slight 
changes in $T_{\rm mag,max}$, as shown in Fig \ref{wind}, led to variations in  
the line intensity and details in the line profile, such as a more 
pronounced redshifted shoulder. The maximum wind temperature
used in the models was $T_{\rm wind,max}=9000$ K and the temperature
of the hot spot was varied from $T_{\rm ring}=5500$ K to 8000 K, without there being much change to the 
line profile. The diskwind starts at 7.81 \rstar, immediately outside the outer magnetospheric radius and
its extension was varied from 30 \rstar \ to 8 \rstar, while the mass outflow rate remained constant 
($\dot{M}_{\rm out}=0.1\dot{M}_{\rm acc}$). We therefore increased the wind density as we 
decreased the outer diskwind radius, although neither significant blueshifted absorption nor emission 
were ever observed in the theoretical \hal \ profiles.
This agrees with the results discussed in \citet{lim10}, where it was shown that the wind contribution
to the line profile is negligible for $\dot{M}_{\rm acc} < 10^{-9}$ \msun ${\rm yr}^{-1}$ owing to 
very low wind densities, which is the case in the present calculations.
The mean observed \hal \ profile is well-reproduced by the model (Fig. \ref{wind}, solid lines), 
and our diskwind component has no visible influence on the line profile at the low
mass-accretion rate of V2129 Oph (dashed lines).
With the parameters of Table \ref{tab_models} we were unable to reproduce the redshifted absorption
component in \hal\ commonly observed in V2129 Oph. We then varied the system inclination 
to mimic a non-aligned magnetospheric and rotation axis. With a lower inclination, on 
the order of $45-50$\degr, a redshifted absorption component was found to be visible, 
but the profile intensity was lower than usually observed.

\section{Conclusions}\label{conclusions}
The magnetospheric accretion scenario predicts that the stellar magnetic field
will interact with the disk and disrupt it when the magnetic pressure overcomes
the ram pressure due to the accretion process. Accretion columns will then be formed
and a hot spot/ring will result at the stellar surface where the accreting
gas hits the star. 
In the past few years, we have seen evidence from observations and numerical
simulations that the star-disk interaction is dynamic and mediated by a stellar 
magnetic field that may vary in time and be composed of different
multipoles that have to be taken into account in the magnetospheric accretion paradigm. 
The magnetic field moments are also generally found to be inclined with respect to the 
rotation axis, creating a non-axisymmetric circumstellar environment, whose observed 
characteristics change on a timescale of days, owing to the system rotation.

We have shown in this paper that the magnetic field configuration proposed by \citet{don11}
for the classical T Tauri star V2129 Oph, which includes dipole and octupole fields inclined with respect to 
the rotation axis, reproduces quite well the photometric and spectroscopic variability observed 
over several rotational cycles of the system. 

We have used the magnetic field configuration obtained from observations in the calculations 
of 3D MHD magnetospheric simulations, whose density, velocity, and scaled temperature structures were
mapped on to a radiative transfer grid to obtain line source functions from which we calculated
the theoretical line profiles.  
The \hal \ and \hbeta \ observed emission profiles vary due to rotation modulation and this variability is
qualitatively well reproduced by the computed theoretical models. The observed profiles are however
wider than the theoretical ones (especially for \hbeta), containing an extended emission in the blue wing
and broader-than-predicted full widths at half maximum. This indicates that additional modeling
is needed to fully describe the accretion flows around CTTSs.

Spectral lines calculated for MHD model accretion flows with pure dipole and 
dipole+octupole fields are similar, since in both cases they form in funnel streams 
that are dominated by the dipole component at the truncation radius. 
However, the correlated variability across the emission line profiles has been most
accurately reproduced by a dipole-only model.
We also calculated emission line profiles with a dipolar and axisymmetric model that included
a magnetosphere and a diskwind, but our diskwind model did not provide any 
significant contribution to the \hal \ line profile of V2129 Oph.

The hot and cold spots of V2129 Oph were found to be almost coincident in phase and at high latitudes
but are expected to be located at different levels, the hot spot at the chromosphere
and the cold spot at the photosphere.  The inferred cold and hot spots 
are able to explain the radial velocity variations in the photospheric and \hei \ (5876 \AA) lines.
The phase coincidence of the hot and cold spots is more accurately explained by
the dipole+octupole models than the dipole-only model, since the octupolar component redirects the funnel flow 
towards the magnetic pole, while in the dipole-only case the magnetic pole and hot spots are 
separate. 

The mass accretion rate of the system is generally around $1.5 \times 10^{-9}$ \msun yr$^{-1}$,
but accretion bursts can occur, as observed, on timescales of days, when the mass accretion
rate is higher by as much as three times its quiescent value. 

The spectroscopic and photometric variabilities observed in V2129 Oph are thus consistent with the 
general predictions of complex magnetospheric accretion models with non-axisymmetric,
multipolar fields.

\hspace{1.5em}


\begin{acknowledgements}
We greatfully acknowledge Christophe Lovis
who obtained part of the HARPS spectra.
SHPA acknowledges support from CAPES, CNPq and Fapemig.
\end{acknowledgements}


\end{document}